\documentclass[a4paper,11pt]{article}

\usepackage{jinstpub} 

\usepackage{natbib}

\usepackage{aas_journal_macros}

\usepackage[inline]{enumitem}

\def\VERSION{Version: 1.0.1 - 2025 Mar 18th; Submitted to JINST: December 20, 2024; Accepted January 6, 2025; Published: 2025 Mar 21st; JINST number {\tt JINST\_057P\_1224}; doi: \url{https://doi.org/10.1088/1748-0221/20/03/P03029}  }

\usepackage{xcolor}
\usepackage{hyperref}
\usepackage{academicons}

\newcommand{\orclogo}{\includegraphics[height=\fontcharht\font`A]{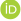}}

\newcommand{\ORCID}[1]{\href{https://orcid.org/#1}{\textcolor{lime}{\orclogo}}}
\title{\boldmath The LSPE-Strip Pointing Reconstruction and Star Tracker}

\author[a,*,**]{Michele Maris,\ORCID{0000-0001-9442-2754}
\note{Corresponding author: M. Maris.}}
\author[b,i]{Maurizio Tomasi,\ORCID{0000-0002-1448-6131}}
\author[b,i]{Matteo Baratto,\ORCID{0009-0000-8702-9591}}
\author[c]{Fabio Paonessa,\ORCID{0000-0001-7660-0502}}
\author[b,i]{Cristian Franceschet,\ORCID{0000-0002-3528-5240}}
\author[a]{Daniele Tavagnacco,\ORCID{0000-0001-7475-9894}}
\author[c]{Oscar Antonio Peverini,\ORCID{0000-0002-7681-6837}}
\author[d]{Fabrizio Villa,\ORCID{0000-0003-1798-861X}}
\author[e,j]{Mario Zannoni,\ORCID{0000-0002-4495-571X}}
\author[b,i]{Marco Bersanelli,\ORCID{0000-0002-7866-948X}}
\author[b,i]{Barbara Caccianiga,\ORCID{0000-0002-8026-7754}}
\author[b,i]{Stefano Mandelli,\ORCID{0000-0001-6737-6422}}
\author[b,i]{Aniello Mennella,\ORCID{0000-0002-5343-6655}}
\author[e]{Federico Nati,\ORCID{0000-0002-8307-5088}}
\author[f]{Stefano Sartor,\ORCID{0000-0002-0012-6848}}
\author[g,h]{Ricardo T. G\'enova-Santos,\ORCID{0000-0001-5479-0034}}
\author[g,h]{Jose A. Rubino-Martin,\ORCID{0000-0001-5289-3021}}
\author[d]{Francesco Cuttaia,\ORCID{0000-0001-6608-5017}}
\author[b,i]{Francesco Cavaliere,}
\author[e,j,k]{Massimo Gervasi,\ORCID{0000-0003-3884-0905}}
\author[a]{and Andrea Zacchei\ORCID{0000-0003-0396-1192}}
\date{\VERSION}

\affiliation[a]{Trieste Astronomical Observatory - Italian National Institute for Astrophysics (INAF-OATs), Via G.B.Tiepolo 11, 34135, Trieste, Italy}
\affiliation[b]{Dipartimento di Fisica ”Aldo Pontremoli”, Università degli Studi di Milano, Via G.Celoria 16, 20133, Milano, Italy}
\affiliation[c]{Istituto di Elettronica e di Ingegneria dell'Informazione e delle Telecomunicazioni (IEIIT), Consiglio Nazionale delle Ricerche (CNR), Corso Duca degli Abruzzi 24, 10129 - Torino, Italy}
\affiliation[d]{INAF – Osservatorio di Astrofisica e Scienza dello Spazio di Bologna, Via Gobetti 93/3, 40129 Bologna, Italy}
\affiliation[e]{Dipartimento di Fisica, Universit\`{a} di Milano - Bicocca, Piazza della Scienza 3, 20126 Milano, Italy}
\affiliation[f]{PCF OAVdA - Osservatorio Astronomico della Regione Autonoma Valle d’Aosta, Lignan, 39 11020 NUS (AO), Italy }
\affiliation[g]{Instituto de Astrof\'{\i}sica de Canarias, E-38200 La Laguna, Tenerife, Canary Islands, Spain}
\affiliation[h]{Departamento de Astrof\'{\i}sica. Universidad de La Laguna (ULL), E-38206 La Laguna, Tenerife, Spain}
\affiliation[i]{INFN--Sezione di Milano, Via Celoria 16, 20133 Milano, Italy}
\affiliation[j]{INFN--Sezione di Milano Bicocca, Piazza della Scienza 3, 20126 Milano, Italy}
\affiliation[k]{CERN, Esplanade des Particules 1,     P.O. Box 1211 Geneva 23, Geneva, Switzerland}
\affiliation[*]{Institute for Fundamental Physics of the Universe (IFPU), Trieste, Italy
Via Beirut, 2, 34151 Trieste (TS), Italy }
\affiliation[**]{ Fondazione ICSC - Centro Nazionale di Ricerca in HPC, Big Data and Quantum Computing Via Magnanelli 2, 40033 Casalecchio di Reno (BO), Italy }

\emailAdd{michele.maris@inaf.it}

\arxivnumber{2501.05604}

\newwrite\ToBeConfirmedFile
\immediate\openout\ToBeConfirmedFile\jobname.tbc
\newcounter{ToBeConfirmed}
\def\ToBeConfirmedNumber{\addtocounter{ToBeConfirmed}{1}TBC-\the\value{ToBeConfirmed}}
\newcommand{\TBC}[1]{\addtocounter{ToBeConfirmed}{1} {\color{blue} #1 $^\mathrm{(TBC-\the\value{ToBeConfirmed})}$}}

\newwrite\ToBeDefinedFile
\immediate\openout\ToBeDefinedFile\jobname.tbd
\newcounter{ToBeDefined}
\def\ToBeDefinedNumber{\addtocounter{ToBeDefined}{1}TBD-\the\value{ToBeDefined}}
\newcommand{\TBD}[1]{\addtocounter{ToBeDefined}{1}{\color{orange} #1 $^\mathrm{(TBD-\the\value{ToBeDefined})}$}}

\def\CAMERA{image plane} 

\def\polQ{Q} 
\def\polU{U} 
\def\polP{P} 
\def\polPideal{P_{\mathrm{unbiased}}} 
\def\polPbiased{P_{\mathrm{biased}}} 
\def\polPbiased{P_{\mathrm{biased}}} 
\def\errorP{\delta P} 
\def\errorPc{\delta P_c} 
\def\gammaPc{\gamma_c} 
\def\gammaPcOne{\gamma_{c,1}} 

\def\Ocam{{\bf \mathcal{O}}_{\mathrm{img}}}

\def\axXcam{{\hat{\mathbf{X}}}_{\mathrm{img}}}
\def\axYcam{{\hat{\mathbf{Y}}}_{\mathrm{img}}}
\def\axZcam{{\hat{\mathbf{Z}}}_{\mathrm{img}}}

\def\eXcam{{\hat{\mathbf{e}}}_{\mathrm{x},\mathrm{img}}}

\def\eZcam{{\hat{\mathbf{e}}}_{\mathrm{z},\mathrm{img}}}

\def\Otel{{\bf \mathcal{O}}_{\mathrm{tel}}}    
\def\axXtel{{\hat{\mathbf{X}}}_{\mathrm{tel}}} 
\def\axYtel{{\hat{\mathbf{Y}}}_{\mathrm{tel}}} 
\def\axZtel{{\hat{\mathbf{Z}}}_{\mathrm{tel}}} 

\def\Otopo{{\bf \mathcal{O}}_{\mathrm{topo}}}

\def\axXtopo{{\hat{\mathbf{X}}}_{\mathrm{topo}}}
\def\axYtopo{{\hat{\mathbf{Y}}}_{\mathrm{topo}}}
\def\axZtopo{{\hat{\mathbf{Z}}}_{\mathrm{topo}}}

\def\cfgVector{\mathbf{\Theta}}                

\def\rollCam{\hat{r}_{\mathrm{img}}}         
\def\panCam{\hat{p}_{\mathrm{img}}}          
\def\tiltCam{\hat{t}_{\mathrm{img}}}         

\def\tiltFork{\hat{t}_{\mathrm{fork}}}       

\def\zVAX{\hat{z}_{\mathrm{VAX}}}            
\def\omegaVAX{\omega_{\mathrm{VAX}}}         
\def\AzVAX{\hat{\mathrm{Az}}_{\mathrm{VAX}}}            

\def\deflP{\hat{\delta}}           

\def\varthetaScan{\vartheta_{\mathrm{scan}}} 

\def\Alt{\widehat{\mathrm{Alt}}}                       
\def\Az{\widehat{\mathrm{Az}}}                         
\def\deltaAlt{\delta \Alt}         
\def\deltaAz{\delta \Az}           
\def\AltZ{\widehat{\mathrm{Alt}}_0}                       
\def\AzZ{\widehat{\mathrm{Az}}_0}                         

\def\RA{\widehat{\mathrm{ra}}}                         
\def\DEC{\widehat{\mathrm{dec}}}                       
\def\deltaRA{\delta \RA}           
\def\deltaDEC{\delta \DEC}         

\def\varthetaZero{\vartheta_{\mathrm{0}}}     
\def\varphiZero{\varphi_{\mathrm{0}}}         

\def\varthetaRef{\vartheta_{\mathrm{ref}}}     
\def\varphiRef{\varphi_{\mathrm{ref}}}         

\def\Long{\widehat{\mathrm{lon}}}                     
\def\Lat{\widehat{\mathrm{lat}}}                      
\def\Height{h_{\mathrm{m}}}                       

\def\PointingSi{{\hat{\mathbf{P}}}^{\mathrm{O}}_i}      
\def\PointingMi{{\hat{\mathbf{P}}}^{\mathrm{M}}_i}      
\def\nsimp{n_{\mathrm{p}}}                              
\def\strdrift{r_{\mathrm{str}}} 
\def\ntheta{n_{\mathrm{c}}} 
\def\nphi{n_{\mathrm{phi}}} 
\def\PointCaver{\mathbf{P}^{\mathrm{0}}_{c}} 
\def\PointCcos{\mathbf{P}^{\mathrm{C}}_{c}} 
\def\PointCsin{\mathbf{P}^{\mathrm{S}}_{c}} 
\def\EncoderAzi{\mathrm{Az}_{\mathrm{enc},i}}  
\def\Pointing{{\hat{\mathbf{P}}}}                
\def\PointingP{{\hat{\mathbf{P}'}}}                

\def\pair{p}                     
\def\Tair{T}                     

\def\rotX{\mathbf{\mathcal{R}}_{\mathrm{x}}} 
\def\rotY{\mathbf{\mathcal{R}}_{\mathrm{y}}} 
\def\rotZ{\mathbf{\mathcal{R}}_{\mathrm{z}}} 

\def\ArotX{\mathbf{\mathcal{R}}_{\mathrm{x}}} 
\def\ArotZ{\mathbf{\mathcal{R}}_{\mathrm{z}}} 

\def\AttitudeMatrix{\mathcal{A}}  
\def\AttitudeMatrixTEL{\mathcal{A}^{\mathrm{(tel)}}}     
\def\AttitudeMatrixVAX{\mathcal{A}^{\mathrm{(V-AXIS)}}}  
\def\AttitudeMatrixGEO{\mathcal{A}^{\mathrm{(geo)}}}  

\def\Nside{N_{\mathrm{side}}} 

\def\ADU{\mathrm{adu}}          
\def\Celsius{^\circ\mathrm{C}}  
\def\Kelvin{\mathrm{K}}  
\def\milliKelvin{\mathrm{mK}}  
\def\microK{\mu\mathrm{K}}      
\def\Km{\mathrm{Km}}            
\def\meter{\mathrm{m}}          
\def\cm{\mathrm{cm}}            
\def\mm{\mathrm{mm}}            
\def\micron{\mu\mathrm{m}}      
\def\GeV{\mathrm{GeV}}          
\def\Hz{\mathrm{Hz}}            
\def\GHz{\mathrm{GHz}}          
\def\sec{\mathrm{s}}            
\def\msec{\mathrm{ms}}          
\def\microsec{\mu\mathrm{s}}    
\def\microAmpere{\mu\mathrm{A}} 
\def\deg{^{\circ}}              
\def\arcmin{\mathrm{arcmin}}    
\def\arcsec{\mathrm{arcsec}}    
\def\pxl{\mathrm{pxl}}          
\def\EncoderAlt{\mathrm{Alt}_{\mathrm{enc}}}                    
\def\EncoderAz{\mathrm{Az}_{\mathrm{enc}}}                      
\def\Planck{{\sc Planck}}        
\def\STRIP{{\sc LSPE-Strip}}    
\def\SWIPE{{\sc LSPE-SWIPE}}
\def\LSPE{{\sc LSPE}}
\def\WMAP{WMAP}

\def\STRIPLSPE{{\sc LSPE-Strip}}

\def\TeideObsLat{+28.3\deg}  

\def\Linux{{\tt Linux}}

\def\AstrometryNet{{\tt Astrometry.net}}  
\def\StripelineJL{{\tt Stripeline.jl}}    

\def\Python{{\tt python}}
\def\PythonTh{{\tt python3}}
\def\astropy{{\tt astropy}}

\def\Cpp{{\tt C++}}

\def\astropy{{\tt AstroPy}}
\def\HEALpix{{\tt HEALpix}}
\def\FITS{{\tt FITS}}

\def\Arduino{{\tt Arduino}}

\def\FWHM{\mathrm{FWHM}}                       
\def\OmegaBeam{\Omega_{\mathrm{beam}}}         
\def\OmegaSource{\Omega_{\mathrm{source}}}     
\def\Tant{T_{\mathrm{ant}}}                    
\def\Tb{T_{\mathrm{b}}}                        

\def\Vlim{V_\mathrm{lim}}           

\def\FOV{\mathrm{FOV}}          
\def\LCAM{L_{\mathrm{cam}}}          
\def\pixelSizeSky{p^{''}}              
\def\pixelSize{p_{\mu}}                

\def\fovAperture{\alpha_{\mathrm{fov}}} 
\def\flength{f}                         
\def\Diam{D}                            
\def\Aopt{A_{\mathrm{opt}}}             

\def\strTelOffset{\delta\mathbf{O}} 

\def\ndetectedstars{n_{*}}

\def\frameStep{\delta t_{\mathrm{frame}}}   
\def\pulseBrigthOn{I_{\mathrm{on}}} 
\def\tpulse{t_{\mathrm{pulse}}}   
\def\pulseWidth{\delta t_{\mathrm{pulse}}}   
\def\pulseShape{I}   
\def\pulseStepLF{\delta t_{\mathrm{pulse}}}   
\def\pulseStepHF{\delta t_{\mathrm{PWN}}}   
\def\pulseStepHFDuty{d_{\mathrm{PWN}}}   
\def\cameraV{V}   
\def\cameraGain{G_{\mathrm{cam}}} 
\def\cameraOffset{O_{\mathrm{cam}}} 
\def\tpps{t_{\mathrm{pps}}}    
\def\tppsZ{t_{\mathrm{pps},0}} 
\def\tppsO{t_{\mathrm{pps},1}} 
\def\tickspps{\tilde{t}_{\mathrm{pps}}}    
\def\ticksppsZ{\tilde{t}_{\mathrm{pps},0}} 
\def\ticksppsO{\tilde{t}_{\mathrm{pps},1}} 
\def\tMPU{t_{\mathrm{mpu}}} 
\def\tMPUPPS{\tilde{t}_{\mathrm{mpu},\mathrm{pps}}}  
\def\delayPulseTrigger{\delta t_{\mathrm{trg}}}  
\def\iFpls{i_{\mathrm{F}}} 
\def\iFplsZ{i_{\mathrm{F},0}} 
\def\iFplsO{i_{\mathrm{F},1}} 
\def\erriFpls{\sigma_{i_{\mathrm{F}}}}     
\def\erriFplsZ{\sigma_{i_{\mathrm{F},0}}}  
\def\erriFplsO{\sigma_{i_{\mathrm{F},1}}}  
\def\tpls{t_{\mathrm{pls}}}      
\def\tplsZ{t_{\mathrm{pls},0}}   
\def\tplsO{t_{\mathrm{pls},1}}   
\def\pulseSigma{\sigma_{\mathrm{pulse}}}  
\def\pulsetPeak{\tilde{t}_{\mathrm{mpu},\mathrm{peak}}}  
\def\pulsePedestal{I_{\mathrm{b}}}  
\def\pulsePeak{I_{\mathrm{peak}}}  
\def\pulsetPeakO{\tilde{t}_{\mathrm{mpu},\mathrm{peak},1}}  
\def\pulsetPeakZ{\tilde{t}_{\mathrm{mpu},\mathrm{peak},0}}  
\def\refSignal{Q_{\mathrm{ref}}}              
\def\ledSignal{Q_{\mathrm{led}}}              
\def\thSignal{\Delta Q_{\mathrm{th}}}         
\def\pulseStrength{\Delta Q_{\mathrm{pls}}}   
\def\frameIndex{i_{\mathrm{f}}}   
\def\naver{N_{\mathrm{aver}}}   
\def\statusWAITPZERO{{\tt WAIT-P0}}
\def\statusWAITNPZERO{{\tt WAIT-NOP0}}
\def\statusWAITPONE{{\tt WAIT-P1}}
\def\statusWAITNPONE{{\tt WAIT-NOP1}}
\def\statusIDLEZERO{{\tt IDLE-0}}
\def\statusPROCESSING{{\tt PROCESSING}}
\def\texp{t_{\mathrm{exp}}}                       
\def\texptotal{t_{\mathrm{exp},\mathrm{tot}}}     

\def\tobs{t_{\mathrm{obs}}}                        
\def\quantumEff{Q_{\mathrm{eff}}}                  

\begin{document}

\abstract{
%
This work is part of a series of papers describing in detail the design and characterization of
the \STRIPLSPE{}, a microwave telescope operating in the Q- and W-bands which is foreseen to be installed at the {\em Observatorio del Teide} in Tenerife. 
%
The paper aims to describe the Pointing Reconstruction Model (PRM) and the prototype Star Tracker, which will be mounted on \STRIPLSPE{}. 
Pointing reconstruction is a crucial step in deriving sky maps of foreground emissions. The PRM will be in charge of integrating the information on the instantaneous attitude provided by the telescope control system, encoded in two control angles, to obtain the actual pointing direction and focal plane orientation of the telescope. The PRM encodes various non-idealities in the telescope setup from eight configuration angles. The Star Tracker, plus the observation of an artificial source installed on a drone and possibly observations of point sources of known positions, will be used to calibrate the configuration angles of the PRM.
%
We study the pointing error produced by incorrectly calibrating configuration angles by comparing surveys with different realizations of systematic pointing errors against the ideal case. In this way, we validated the required $\approx 1\;\mathrm{arcmin}$ maximum systematic pointing error in the \STRIPLSPE{} survey as the worst effect of the pointing error, in this case, is two orders of magnitude below the instrumental sensitivity.
After a description of the main structure and operations of the Start Tracker, we present the results of a campaign of actual sky observations carried out on a prototype of the Star Tracker aimed at assessing the final Star Tracker accuracy. From the point of view of performance, the Star Tracker prototype fully represents the final Star Tracker, the main differences being related to several implementation details. The results show a Star Tracker RMS accuracy is $\approx 3\;\mathrm{arcsec}$ while the systematic error is below $10\;\mathrm{arcsec}$. 
From those results, we analyzed the problem of reconstructing the PRM configuration angles by simulating a calibration survey. Given the need to intercalibrate the offset of the Start Tracker pointing direction with respect to the focal plane pointing direction, we simulated two possible intercalibration strategies: one by simulating intercalibration with the use of observations of planets, the second by observing a drone carrying an optical beacon and a radio beacon. In the first case, the accuracy is determined by the level of $1/f$ instrumental plus atmospheric noise, determining the S/N by which the planet can be observed. A very conservative S/N=10 case and a more likely S/N=50 case have been considered, allowing for an intercalibration accuracy respectively of $1\;\mathrm{arcmin}$ and $1/3\;\mathrm{arcmin}$. In the second case, the most important source of error is the correct evaluation of the parallaxes between the telescope and the Star Tracker. Our analysis shows that the intercalibration accuracy will be between $0.25\;\mathrm{arcmin}$ and $1\;\mathrm{arcmin}$ in the worst cases.
%
}


\footnotetext[0]{\VERSION}

\maketitle

\section{Introduction}\label{sec:introduction}

The polarization of the Cosmic Microwave Background (CMB) encodes a wealth of information about the early stages of the evolution of the universe. The E-mode pattern is sensitive to the optical depth of the ionized medium, so its amplitude provides a powerful indication of the re-ionization era 
(see, e.g. \cite{2016ASSL..423.....M}). Furthermore, accurate measurements of the E-mode are vital to breaking the degeneracies between the cosmological parameters extracted from the temperature anisotropies. Although the E-mode power spectrum has been measured with good accuracy over a wide multipole range by the \WMAP\ and \Planck\ space missions 
\cite{2013ApJS..208...19H,2016A&A...596A.108P},  current observations are still limited by instrumental and astrophysical uncertainties rather than by cosmic variance, particularly at large angular scales.

An even more exciting scientific motivation driving CMB polarization experiments is the quest for the B-mode component. If detected, B-modes would provide strong evidence of primordial gravitational waves produced during an inflation era that occurred in the very early universe
\cite{1997PhRvL..78.2054,1997PhRvL..78.2058K,1997PhRvD..56..596H},
possibly at energies of order $\sim 10^{15}\;\GeV$. The unique opportunity to probe physics at such uncharted conditions motivates an extended worldwide effort to measure CMB polarization to high precision, with experiments operating from ground, balloon, and space 
(see, eg., \cite{2019JCAP...02..056A,2022JCAP...04..034H,2022ApJ...926...54A,2023PTEP.2023d2F01L}
and references therein).

The experimental challenges are enormous because the B-mode signal is highly faint (well below $1\mu K$) and overwhelmed by diffuse foreground emission. Observations must be conducted over a wide frequency range to disentangle the primordial signal from polarized synchrotron and dust emission originating within our Galaxy, ensuring that ground-based instruments properly avoid spectral bands hampered by atmospheric opacity. At the same time, systematic effects must be controlled at extreme levels, which requires detailed characterization of instrumental features such as noise properties, beams, thermal stability, and pointing reconstruction. 
The current upper limits on the tensor-to-scalar ratio parameter, characterizing the relative amplitude of the B-mode component, come from a combination of BICEP/Keck 
\cite{2021PhRvL.127o1301A}
and \Planck/PR4 
\cite{2020A&A...643A..42P}
data, yielding $r < 0.032$ at 95\% confidence level 
\cite{2022PhRvD.105h3524T}.

One such project, the Large Scale Polarization Experiment (\LSPE), combines ground-based and balloon-borne observations to reach wide frequency coverage while optimizing instrumental characteristics in the presence of atmospheric effects. 
 %
\cite{2021JCAP...08..008A}.
The \SWIPE{} instrument will be flown in an Arctic stratospheric flight and will measure at frequency bands centered at 145, 210, and 240 GHz with an array of multimode bolometers. At lower frequencies, \STRIPLSPE{} will observe the sky with an array of 49 coherent polarimeters at 43\,GHz, providing key information on the Galactic synchrotron, and six channels at 95\,GHz mainly as atmospheric monitor. Both \STRIPLSPE{} and \SWIPE{} will cover nearly the same area of the Northern Hemisphere, corresponding to about $25\%$ of the whole sky. The baseline balloon flight of \SWIPE{} is a 2-week polar trajectory around the North Pole during the arctic night. Instead, the \STRIPLSPE{} telescope will be installed at the Teide Observatory, Tenerife, and will integrate for 2 years with a spinning scanning strategy. The combined data set is expected to improve the current B-mode limits, reaching $r < 0.015$ at 95\% 
\cite{2023MNRAS.519.3383R}, 
carried out from the same site, enabling us to improve our understanding of polarized synchrotron emission, a crucial component for all CMB polarization experiments.

This work is part of a series of articles that describe in detail the design and characterization of the \STRIPLSPE{} instrument, some of which are published \cite{lspebeams2022,2022JInst..17P1029F,2022JInst..17P6042P,Genova-Santos:etal:2023},
while others are in preparation. Here, we discuss the pointing system of the telescope and our strategy for extracting the information needed to support the data analysis. Ensuring adequate control of the telescope attitude over the whole observing campaign and a quantitative estimate of the related uncertainties is necessary and challenging. In the following, we present the pointing model of \STRIPLSPE{} and the strategy we devised to reconstruct the pointing direction of the detectors based on a dedicated Star Tracker and on attitude information routinely acquired as part of the housekeeping.

The paper is structured as follows. 
In Sect.~\ref{sec:generalities}, we provide a synthetic overview of the \STRIPLSPE{} telescope 
and discuss the basic requirements for pointing accuracy
 %
An analysis of the pointing reconstruction is the subject of Sect.~\ref{sec:PRM}: here we start from the formal definition of the reference frames we adopted (Sect.~\ref{sec:ref:frames}), then we present an analytical model for the telescope attitude (Sect.~\ref{sec:method}), we explain the metric we used to evaluate the pointing accuracy
(Sect.~\ref{sec:metrics:pointin:accuracy}), and we discuss the expected residual systematic effects due to pointing uncertainties (Sect.~\ref{sec:prm:configuration:errors}).
In Sect.~\ref{sec:str} we illustrate the design of the \STRIPLSPE\ Star Tracker and describe its operating modes: we  present the bread-board model developed to verify the Star Tracker  performance (Sect.~\ref{sec:str:implementation}), we provide the details of the time synchronization system (Sect.~\ref{sec:str:sync}), we present the main results of our breadboard test campaign  (Sect.~\ref{sec:str:testing}); we then use those results to predict the level of attainable pointing accuracy for \STRIPLSPE{} by calibrating the PRM with the STR (Sect.~\ref{sec:str:prm}) and we discuss the accuracy of the Star Tracker / Telescope pointing intercalibration 
(Sect.~\ref{sec:prm:intercalibration}).
Finally, the conclusions and an outline of future planned activities are given in Sect.~\ref{sec:conclusions}.

\section{\STRIPLSPE{} Instrument and pointing framework}\label{sec:generalities}


\begin{figure}
\centering
\includegraphics[width=0.67\textwidth]{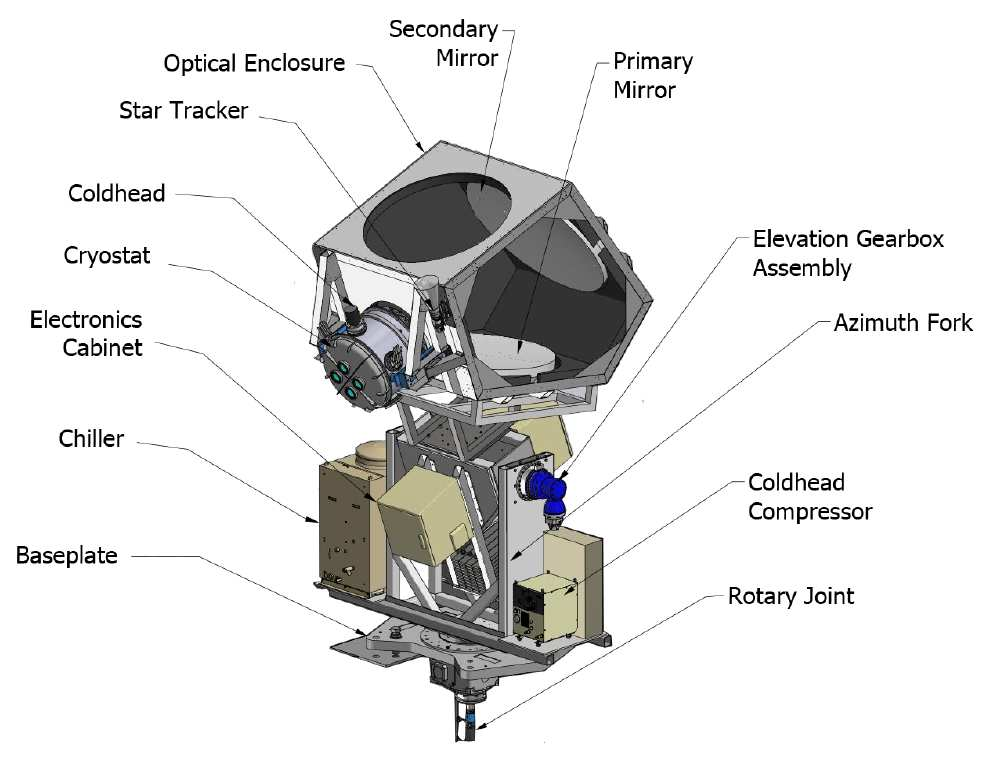}
\caption{\label{fig:strip:scheme}
Structure of the \STRIP{} telescope. The telescope rotates around its main vertical axis, hosting a rotary joint that allows electric power transmission and services. The gearbox of the vertical axis is connected to the ground through a base plate, which can be aligned with the zenith and the local meridian. The vertical axis is joined to a frame hosting the chiller, the cold head compressor, and the azimuth fork. The telescope frame, holding the cryostat with the focal plane, the optics, and the electronic cabinets, is connected at the elevation axis, allowing it to change its elevation.  
}
\end{figure}

\begin{figure}
\centering
\includegraphics[width=0.67\textwidth]{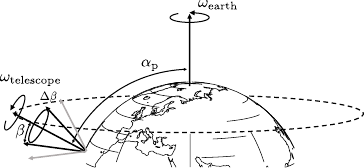}
\caption{\label{fig:strip:functional:scanningStrategy}
The way \STRIP{} observes the sky. The boresight direction of the telescope is kept at a constant altitude of 70$^\circ$, and the Azimuth motor spins regularly around the vertical axis once every minute ($\omega_\text{telescope} = 1\,\mathrm{rpm}$). Coupled with the daily rotation of Earth ($\omega_\text{earth}$), this lets the Strip detectors observe $\sim$30\,\% of the sky after one day.}
\end{figure}

The objective of \STRIP{} is to characterize the polarized sky emission at low frequencies, which is used in conjunction with the measurements produced by \SWIPE{} to separate foregrounds from the CMB signal. An overview of the \STRIP{} telescope is shown in
Fig.~\ref{fig:strip:scheme}. 

The telescope is mounted on an alt-azimuth mount, hosted in a telescope enclosure that minimizes stray light from the ground. The telescope optical system is based on a dual-reflector off-axis design with a parabolic primary with apertures roughly $1.5\;\meter$ 
and a hyperbolic secondary 
of aperture $1.7\;\meter$. 
An array of forty-nine corrugated feed horns at Q and six at W band couple the telescope with a set of polarimeters based on HEMT cryogenic amplifiers cooled to $20\;\Kelvin$. The horns interface the telescope, resulting in pointing directions in the sky within an angular range of $\pm 5\deg$ from boresight 
The \STRIP{} baseline scanning scheme foresees a continuous rotation, made possible by a rotary joint mounted at the bottom of the mechanical system, which ensures transmission of the power and the scientific signal during the observations.

The \STRIP{} and \SWIPE{} instruments cover roughly the same portion of the sky ($\sim 25\%$ of the whole sky) with widely different scanning strategies. \SWIPE{} will be operated at stratospheric altitudes and will cover the required sky region by spinning the instrument while moving around the North Pole. The \STRIP{} telescope, to be installed at the Teide Observatory, Tenerife (Lat = $\TeideObsLat$), will spin around a vertical axis at a constant rotation speed of 1\,rpm, and pointing at a nominal angular distance from the Zenith  $\beta = 20^\circ$, as shown in Fig.~\ref{fig:strip:functional:scanningStrategy}. By exploiting the Earth's daily rotation, the projected scan circle will cover the required angular band with a height of roughly 40 degrees. More precisely, the fraction of  sky observed by  \STRIP{} can be computed from the difference between the solid angles of two spherical caps around the North Pole, one with radius $90^\circ - \phi_\text{topo} + 20^\circ$, and the other with radius $90^\circ - \phi_\text{topo} - 20^\circ$:

\begin{equation}
f_\text{sky} \approx \frac{2\pi\bigl(1 - \cos(90^\circ - \phi_\text{topo} + 20^\circ)\bigr) - 2\pi\bigl(1 - \cos(90^\circ - \phi_\text{topo} - 20^\circ)\bigr)}{4\pi} \approx 30\,\%,
\end{equation}

\noindent
where $\phi_\text{topo} \approx 28^\circ\,\text{N}$ is the latitude of the site.
As explained in \cite{thelspecollaboration2020large}, this value of $\beta$ will ensure that \STRIP{} and \SWIPE{} observe roughly the same horizontal strip in the Northern Hemisphere.   

\begin{figure}
\centering
\includegraphics[width=\textwidth]{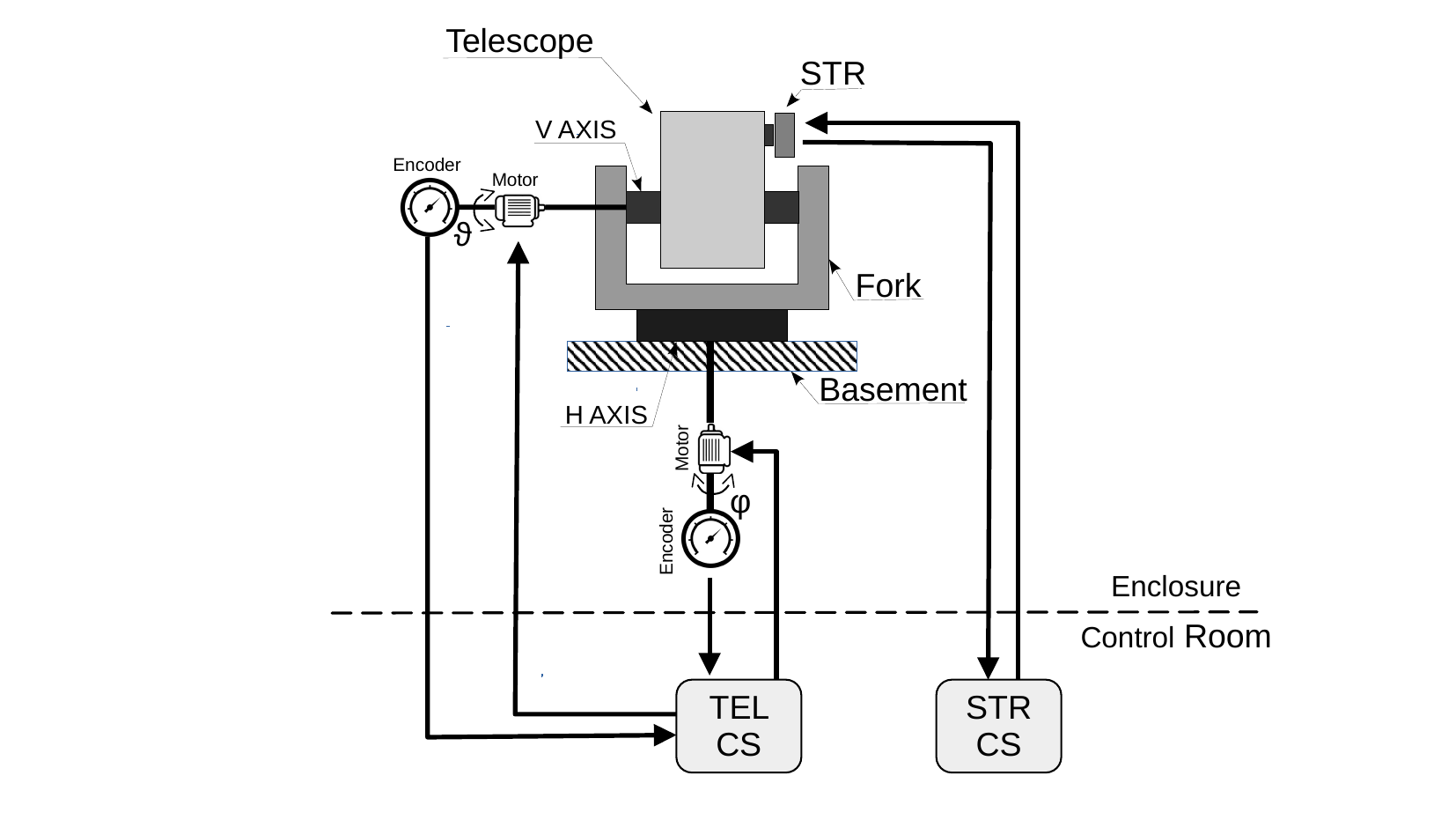}
\caption{\label{fig:strip:functional:units}
Main functional units of \STRIP{}.
In the real telescope, encoders are not mounted on the motor shafts but onto the H-AXIS and V-AXIS shafts.
}
\end{figure}

In Figure~\ref{fig:strip:functional:units} we illustrate the relevant functional elements to describe the \STRIP{} pointing system. The scheme highlights the relevant telescope elements without any reference to their mechanical properties, with the sole purpose of defining valuable quantities.
The alt-azimuth mount is shown consisting of a vertical axis (V-AXIS) around a {\it Fork} and a horizontal axis (H-AXIS) around the {\it Basement}. 
Each axis is equipped with its own motor and absolute encoder. The two angles associated with these axes are the {\em control angles} $\vartheta$ and $\varphi$.
The {\em Telescope Control Station} (TEL CS) controls the telescope in the Control Room. The pointing direction toward which the telescope is aimed can be monitored at night through the Star Tracker (STR), which is mounted on the telescope. 
The Start Tracker Control Station (STR CS) controls the STR, which is also located in the Control Room.
Communication between the units in the telescope Enclosure and the Control Room is done through an internal Ethernet network (not shown in the figure).
For the sake of simplicity, the plot encoders are shown as mounted on the motor shafts; in reality, encoders are mounted directly on the H-AXIS and V-AXIS shafts\footnote{Motors have their encoders, but they are not used to determine the rotation angles of the telescope.}.
Finally, all the telescope units are synchronized through an internal Master Clock (not shown in the figure). The Master Clock is synchronized with UTC through a dedicated GPS receiver.

Proper reconstruction of the pointing information associated with each sample acquired by the 55 detectors is crucial since this information is required to combine the time-ordered data into sky maps. The Strip data reduction pipeline needs to integrate multiple observations of the same regions of the sky to produce an estimate of the polarized emission over the fraction of sky covered by the scanning strategy (\emph{map-making}), and any inaccuracy will lead to systematic errors that propagate to the component separation stage and the scientific data reduction.

The primary source of information in determining pointing are the encoders and the calibration of the pointing model performed by using the Star Tracker, so to characterize the extent of systematic errors in pointing reconstruction, one has to assess how the accuracy of these devices affects the estimated orientation of the \STRIP{} optical elements.

An essential outcome of this work will be to set well-motivated pointing requirements for STRIP. As a helpful reference, the \Planck{}/LFI 44\,GHz channel worked at similar angular resolution and frequency as the STRIP array. In the case of Planck, the pointing requirement was $30\,\arcsec$, with a goal of $15\,\arcsec$. However, those limits were driven by higher frequency HFI channels. As a conservative starting point, we initially adopted the exact requirements for STRIP as in Planck. Still, they were never tested using a dedicated end-to-end simulation of the Strip instrument. As we will discuss in Sect.~\ref{sec:prm:configuration:errors}, as a result of our analysis, we can relax these requirements based on our pointing reconstruction model and the propagation of the main effects on the STRIP  performance.

\section{Pointing Reconstruction}\label{sec:PRM}


\textit{Pointing reconstruction} is the process of deriving the actual pointing direction of an instrument using the information provided by the system data acquisition. In the case of \STRIP{}, the primary information will be provided by the TCS, and it consists of the control angles $\vartheta$, $\varphi$ as a function of time. Additional key information will be available from the telescope design, data from sensors not controlled by the TCS, and dedicated measurement campaigns. This information must be considered throughout the observing campaign since slight changes and slow drift may occur.
Our Pointing Reconstruction Model (PRM) describes in detail how the components of the \STRIP{} structure determine the nominal pointing direction of the telescope and provides a path to include the effect of external perturbations and non-idealities (such as flexures, etc.) on the instrument pointing.
This section assumes that the telescope has a stable and well-defined geometry with ideal stiffness (i.e., flexures are not considered). In these conditions, the control angles can determine the attitude through a transfer matrix parameterized by a set of {\em configuration angles}, which are assumed to be measured during the commissioning phase at the observing site.  

\subsection{Reference frames}\label{sec:ref:frames}

To determine the pointing direction $\Pointing$ for an observer located at a given place and time, defining a {\em Local Topocentric Reference Frame} (LTRF) is helpful.  
The LTRF would be equivalent to the telescope Alt-Az reference frame for an ideal telescope.
However, all the manufacturing uncertainties of the telescope and its mount make those two reference frames slightly
different. The scope of the PRM is to provide a framework to describe, measure, and account for such differences.

In defining the LTRF, we follow the usual astronomical convention for which
has origin $\Otopo$ in the observer location, X axis $\axXtopo$ oriented toward South, Y axis $\axYtopo$ oriented toward East, and Z axis $\axZtopo$ defined by the local Zenith.
We assumed that $\Otopo$ is at the center of the telescope basement, and it is identified by its longitude $\Long$, latitude $\Lat$, and height $\Height$ above sea level. 
The model for the shape of the Earth is the WGS84 ellipsoid, in agreement with the convention used by the \astropy\ package\footnote{\url{http://www.astropy.org}}.
For the Alt-Az coordinates, we follow the convention of the \astropy\ package for which
\emph{Altitude} ($\Alt$) is the angle between $\Pointing$ and the local horizontal plane. It is favorable for pointing directions above the horizons\footnote{Another name for \emph{altitude} is \emph{elevation angle}, so this reference frame is also known as the Azimuth-Elevation system.}. 
\emph{Azimuth} ($\Az$) is the angle between the projection of $\Pointing$ onto the local horizontal plane and the North direction; it is oriented East of North (i.e., $N=0$, $E=90^\circ$). 
Having defined $\Az$ as positive for a clockwise rotation, the corresponding reference frame is no longer right-handed. To have a right-handed reference frame, we define the control angles  
\begin{equation}\label{eq:control:altaz}
\begin{array}{ccc}
\vartheta & = & 90\deg-\Alt,\\
\varphi& = & 180\deg-\Az;
\end{array}
\end{equation}

\noindent
in this way, $\vartheta$ is the Zenithal distance of $\Pointing$ in the Topocentric reference frame,
while $\varphi$ is anticlockwise and it is oriented South of East ($0\deg=\mathrm{S}$, $90\deg=\mathrm{E}$).

To convert $(\vartheta,\varphi)$ into a pointing direction in the LTRF, $\hat{\mathbf{P}}^{\mathrm{(geo)}}$, the model of the telescope and its Alt-Az mount shown in Fig.~\ref{fig:strip:functional:units} is used.
A basement holds a vertical axis (V-AXIS), which allows the rotation of $\Az$;    
a fork mounted on the top of the V-AXIS holds the Horizontal Axis (H-AXIS), which rotates the telescope in $\Alt$.
Ideally, the H-AXIS is normal to the V-AXIS, and the V-AXIS is aligned with the local (Topocentric) zenith.
The angles $\vartheta$ and $\varphi$ represent rotation along the H-AXIS and the V-AXIS, respectively.

To identify the telescope area collecting electromagnetic radiation from the sky in the pointing direction, we introduce the concept of 
\textit{\CAMERA}.
This definition includes the microwave focal plane and an optical camera like the Star Tracker. In Fig.~\ref{fig:strip:functional:units}, it is represented by the light gray hexagonal region. In our simplified scheme, it is assumed to be located at the bottom of the telescope tube, aimed at the sky.

\begin{figure}
\centering
\begin{tabular}{ccc}
\multicolumn{1}{l}{a.)} & 
\multicolumn{1}{l}{b.)} & 
\multicolumn{1}{l}{c.)} \\
\includegraphics[width=0.3\textwidth]{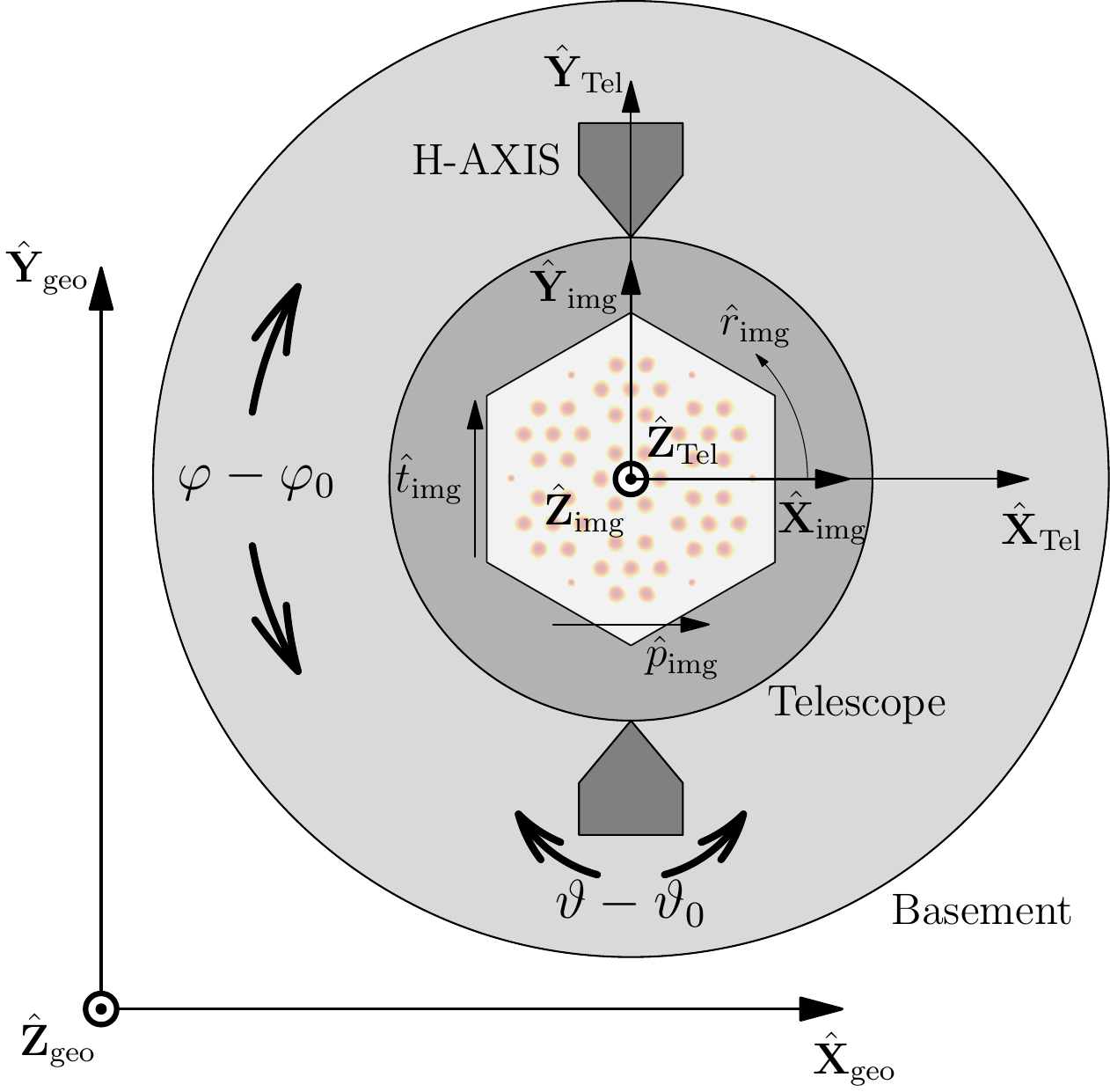}
&
\includegraphics[width=0.3\textwidth]{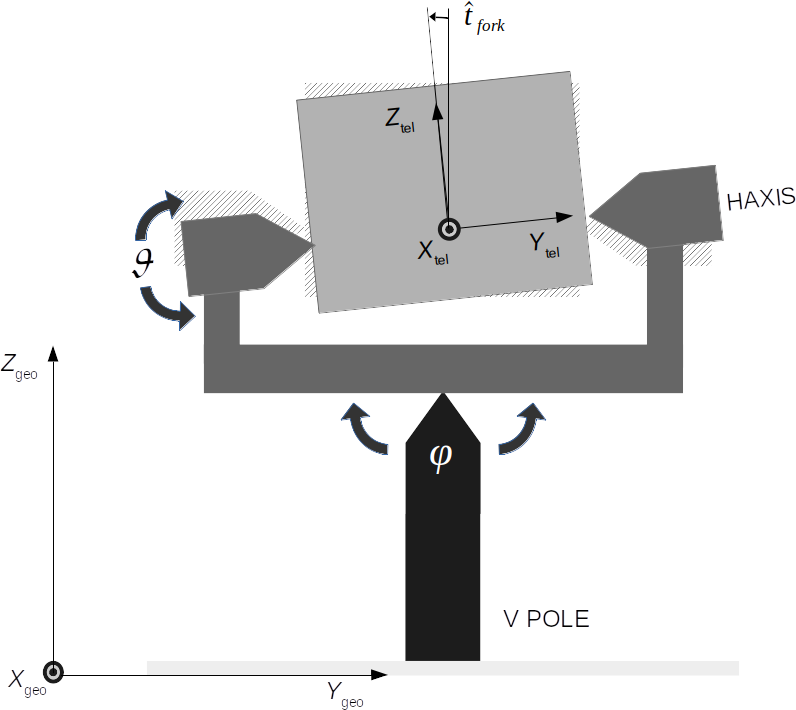}
&
\includegraphics[width=0.3\textwidth]{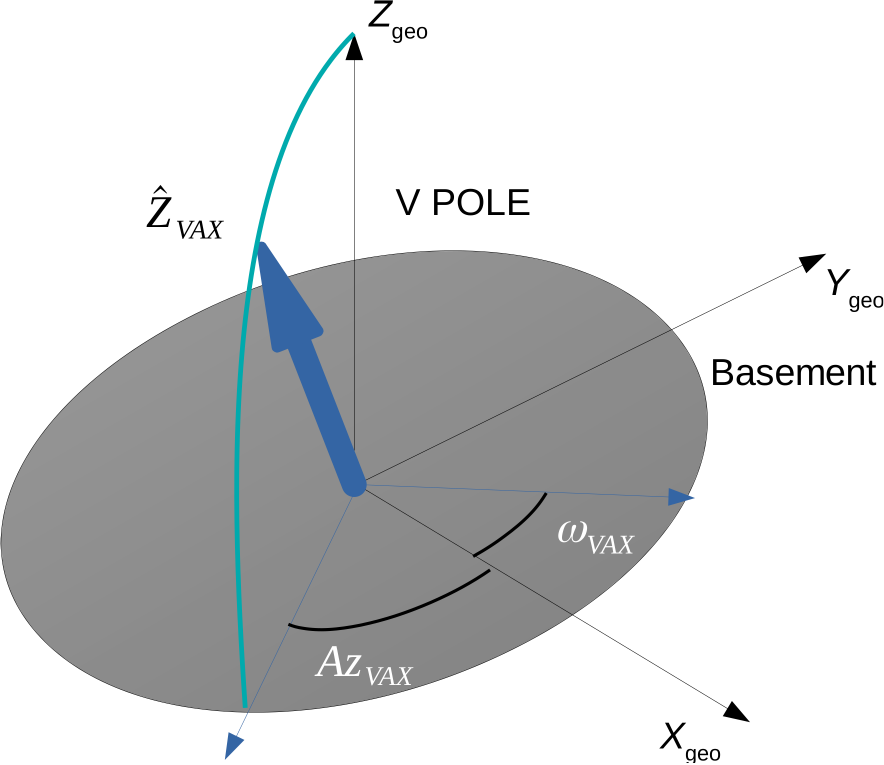}
\end{tabular}
\caption{\label{fig:frames:angles}
Reference frames and angles in the pointing reconstruction model.
Frame a.) the \CAMERA\ reference frame, the telescope reference frame, and the local topocentric reference frame. The telescope is oriented to the zenith, and the axes of the three reference frames are parallel. With the ideal configuration of the telescope, in this case, the axes and origins would overlap, but to avoid confusion when needed, they are shifted in the figure. The figure also shows the \emph{\CAMERA}\ as a light grey hexagon, the telescope body, the H-AXIS, and the basement, as well as the $\varphi$, $\varphiZero$, $\vartheta$, $\varthetaZero$, 
$\rollCam$, $\panCam$ and $\tiltCam$ angles.
Frame b.) shows the $\tiltFork$ angle. 
Frame c.) shows the V-AXIS and the wobble angles $\zVAX$, $\mathrm{Az}_{\mathrm{VAX}}$ and 
$\omegaVAX$.
}
\end{figure}

In Fig.~\ref{fig:frames:angles}, we show graphically the various reference frames used in this analysis and their mutual relationships. Figure~\ref{fig:frames:angles}a  shows the \textit{\CAMERA\ reference frame} and the \textit{telescope reference frame} (the two main internal reference frames used in this work) and their relation with the LTRF. Here, we assume that the telescope is pointing at the Zenith, and the axes of all reference frames are parallel. All the reference frames are right-handed, with positive angles defined for anti-clockwise rotations. 
The origin $\Ocam$ of the \CAMERA\ reference frame is at the center of the hexagon representing the \STRIP\ focal plane. The X-axis $\axXcam$ is aligned with the major axis of the \CAMERA, the Y-axis $\axYcam$  is aligned with the minor axis, and the Z-axis $\axZcam$ is normal to the \CAMERA\ and exits from the telescope aperture. 
The origin of the \CAMERA\ is located at the center of the hexagon.
The origin $\Otel$ of the telescope reference frame is centered on the telescope. The X-axis $\axXtel$ is normal to the H-AXIS, the Y-axis $\axYtel$  is parallel to the H-AXIS, and the Z-axis $\axZtel$ exits from the telescope aperture. 

It is important to stress how this model simplifies the more complex optical system, which includes a larger number of optical components and reference frames. However, if needed, the model described here is sufficiently modular to integrate this extra complexity.

\subsection{The pointing reconstruction model}\label{sec:method}

The pointing direction $\Pointing$ of the system illustrated in Fig.~\ref{fig:strip:functional:units} and Fig.~\ref{fig:frames:angles} is defined by the direction toward which $\axZcam$ is aimed in the LTRF for given values of the control angles $(\vartheta, \varphi)$.
The definition of a PRM is then a matter of computing an attitude operator $\AttitudeMatrix(\vartheta,\varphi, \cfgVector)$ describing the projection of the $\axXcam$, $\axYcam$, $\axZcam$ axes in the LTRF, where $\cfgVector$ is a vector of configuration parameters.
We derive the PRM by applying a chain of orthogonal rotations.

Following Fig.~\ref{fig:frames:angles} we start defining the orientation of the \CAMERA\ relative to the telescope described by the Tait–Bryan angles: 
{\em roll} is an anticlockwise rotation of an angle $\rollCam$ around the $\axZcam$,
{\em pan} is an anticlockwise rotation of an angle $\panCam$ around the $\axYcam$  \footnote{Also named {\em Pitch}.},
{\em tilt} is an anticlockwise rotation of an angle $\tiltCam$ around the $\axXcam$ \footnote{Also named {\em Yaw}.}.
With this convention, a roll rotates the image seen by the \CAMERA\ around its center, while a small pan (or tilt) shifts the image along the \CAMERA\ X (or Y) axis; the change of coordinated from 
the \CAMERA\ to the telescope is described by the operator $\AttitudeMatrixTEL(\tiltCam,\panCam,\rollCam)$.
The projection of $\hat{\mathbf{P}}^{\mathrm{(tel)}}$ into the LTRF is a function of the $\vartheta$ and $\varphi$ control angles. 
 
In the ideal telescope, $\vartheta$ describes a rotation around the $Y$ axis, while $\varphi$ a rotation around the $Z$ axis, 
In reality, one should take into account that the zero points of the devices determining
 $\vartheta$ and $\varphi$ will not perfectly align with the LTRF. For this reason, two zero point angles are introduced: $\varthetaZero$ and $\varphiZero$.

Furthermore, in a real telescope, the H-AXIS and the V-AXIS are not perfectly orthogonal. This is accounted for by the $\tiltFork$ angle, defined in the central frame of Fig.~\ref{fig:frames:angles} and describes a rotation about the X axis.
So the $\AttitudeMatrixVAX$ describing the change of reference frame from the telescope to the V-AXIS
depends on the two control angles $\vartheta$ and $\varphi$ and three configuration angles $\varthetaZero$ and $\varphiZero$ and $\tiltFork$.
 
In addition, one has to consider that V-AXIS could be slightly displaced with respect to the local Zenith, as shown in the right panel of Fig.~\ref{fig:frames:angles}. The corresponding operator $\AttitudeMatrixVAX$ describing the transformation from V-AXIS to the geographic reference plane will depend on two further configuration angles, called {\em wobble angles} $(\zVAX, \omegaVAX)$. Here, $\zVAX$ is the displacement with respect to the V-AXIS, while $\omegaVAX$ is the Azimuth of the ascending node, which is related to the azimuth of the V-AXIS $\AzVAX$ by $\omegaVAX=90\deg+\AzVAX$. 
 
 In conclusion, the attitude operator corresponding to a vector of configuration angles $$\cfgVector = (\omegaVAX,\zVAX,\varphiZero,\tiltFork,\varthetaZero,\tiltCam,\panCam,\rollCam)$$ is obtained by composing the following chain of rotations:

\begin{equation}\label{eq:AttitudeMatrix}
\begin{array}{ccc}
\AttitudeMatrix(\vartheta,\varphi,\cfgVector)  &=& \AttitudeMatrixGEO \AttitudeMatrixVAX(\vartheta,\varphi) \AttitudeMatrixTEL \\
&&\\
\AttitudeMatrixTEL &=& \rotX(\tiltCam) \rotY(\panCam) \rotZ(\rollCam) \eZcam; \\
&&\\
\AttitudeMatrixVAX(\vartheta,\varphi) & = & \rotZ(\varphi-\varphiZero) \rotX(\tiltFork) \rotY(\vartheta-\varthetaZero); \\
&&\\
\AttitudeMatrixGEO & = & \rotZ(\omegaVAX) \ArotX(\zVAX)  \ArotZ(-\omegaVAX).
\end{array}
\end{equation}

\noindent
where rotations are expressed in the form of the usual right-handed rotation matrices\footnote{
For an anticlockwise rotation angle $\alpha$ 
\begin{equation}\label{eq:intrinsic:rotations}
  \begin{array}{ccccccccc}
  \rotX(\alpha) & = & 
     \left( 
       \begin{array}{ccc}
       1 & 0 & 0\\
       0 & \cos \alpha & -\sin \alpha\\
       0 & \sin \alpha &  \cos \alpha
       \end{array}
     \right),
  &
  \rotY(\alpha) & = & 
     \left(
       \begin{array}{ccc}
       \cos \alpha & 0 &  \sin \alpha\\
       0 & 1 & 0\\
       -\sin \alpha & 0 &  \cos \alpha\\
       \end{array}
     \right),
  &
  \rotZ(\alpha) & = & 
     \left(
       \begin{array}{ccc}
       \cos \alpha &  -\sin \alpha & 0 \\
       \sin \alpha  &  \cos \alpha & 0\\
       0 & 0 & 1 \\
       \end{array}
     \right).
  \\
  \end{array}
\end{equation}
}.
The pointing direction in the LTRF is then $\hat{\mathbf{P}}^{(\mathrm{geo})} = \AttitudeMatrix(\vartheta,\varphi,\cfgVector)\eZcam$, while the orientation of the \CAMERA\ is given by  $\hat{\mathbf{O}}^{(\mathrm{geo})} = \AttitudeMatrix(\vartheta,\varphi,\cfgVector)\eXcam$.


From Alt-Az coordinates, the pointing direction has to be converted to the astrometric reference frame, which for \STRIPLSPE\ is the J2000 ICRF. This is a well-known application of positional astronomy\footnote{This is implemented in standard astrometric packages, such as  \astropy\,  which has been used for this work and is well described in standard texbooks}. Therefore, we do not enter details for this step.

\subsection{Metrics for pointing accuracy}\label{sec:metrics:pointin:accuracy}

Various metrics can be used to quantify the pointing accuracy, i.e., the difference
between the actual pointing direction $\PointingP$ and the nominal pointing direction
$\Pointing$. The simplest quantity is the deflection angle:
\begin{equation}\label{eq:deflP}
  \deflP = \arccos \PointingP\cdot\Pointing,
\end{equation}
which is always positive and does not depend on the reference frame.

For the Alt-Az reference frame, $\PointingP$ can be decomposed in $\Alt'$ and $\Az'$; the corresponding errors are:
\begin{align}
\label{eq:error:alt}
\deltaAlt &= \Alt' - \Alt\\
\label{eq:error:az}
\deltaAz &= \Az' - \Az.
\end{align}

\noindent
For small differences in $\Alt$ and $\Az$,
\begin{equation}\label{eq:deflP:small}
  \deflP^2 \approx \deltaAlt^2 + \deltaAz^2 \cos^2\Alt.
\end{equation}
From this, if $\deltaAlt=0$, then $\deflP = |\deltaAz| \cos \Alt $ so that a significant $\Az$ error
near the Zenith corresponds to a small pointing error. For example, in a nominal scan of \STRIP{} with $\Alt=70\deg$ ($\varthetaScan=20\deg$), the effect of $\deltaAz$ on the deflection is scaled by a factor $\sim 0.34$.
Similarly, it is possible to define metrics for pointing accuracy in the equatorial reference frame
$\deltaRA$ and $\deltaDEC$ by replacing in
Eq.~\eqref{eq:error:alt}, Eq.~\eqref{eq:error:az}, and Eq.~\eqref{eq:deflP:small} $\Alt$ with $\DEC$ and $\Az$ with $\RA$.


\subsection{Effect of uncertainties in the PRM configuration angles}\label{sec:prm:configuration:errors}

We have run simulations to estimate the impact of uncertainties in our knowledge of the configuration angles using the library \StripelineJL\footnote{\url{https://github.com/lspestrip/Stripeline.jl}}. Our simulations compute the attitude operator in Eq.~\eqref{eq:AttitudeMatrix}, given the control and configuration angles with some bias. Then, they project the direction of observation into the Equatorial Celestial reference frame as a function of the position of the observation site and the epoch (see Sec.~\ref{sec:method}). Finally, we generate simulated Time-Ordered Data (TOD) at the nominal sampling frequency using the value of the zenith control angle $\varthetaScan = 20\deg$, as defined in the baseline scanning strategy. To ease the interpretation of the data, we do not add instrument noise to the TODs so that the only perturbations to the ideal cases are the bias in the configuration angles.
 %
To simulate beam convolution, the input maps used by the simulator to generate a TOD are convolved with a Gaussian beam with $\FWHM = 20~\arcmin$, i.e., the nominal mean FWHM at $43~\GHz$ for \STRIPLSPE{} beams.

\begin{figure}[!h]
    \centering
    \includegraphics[width=\textwidth]{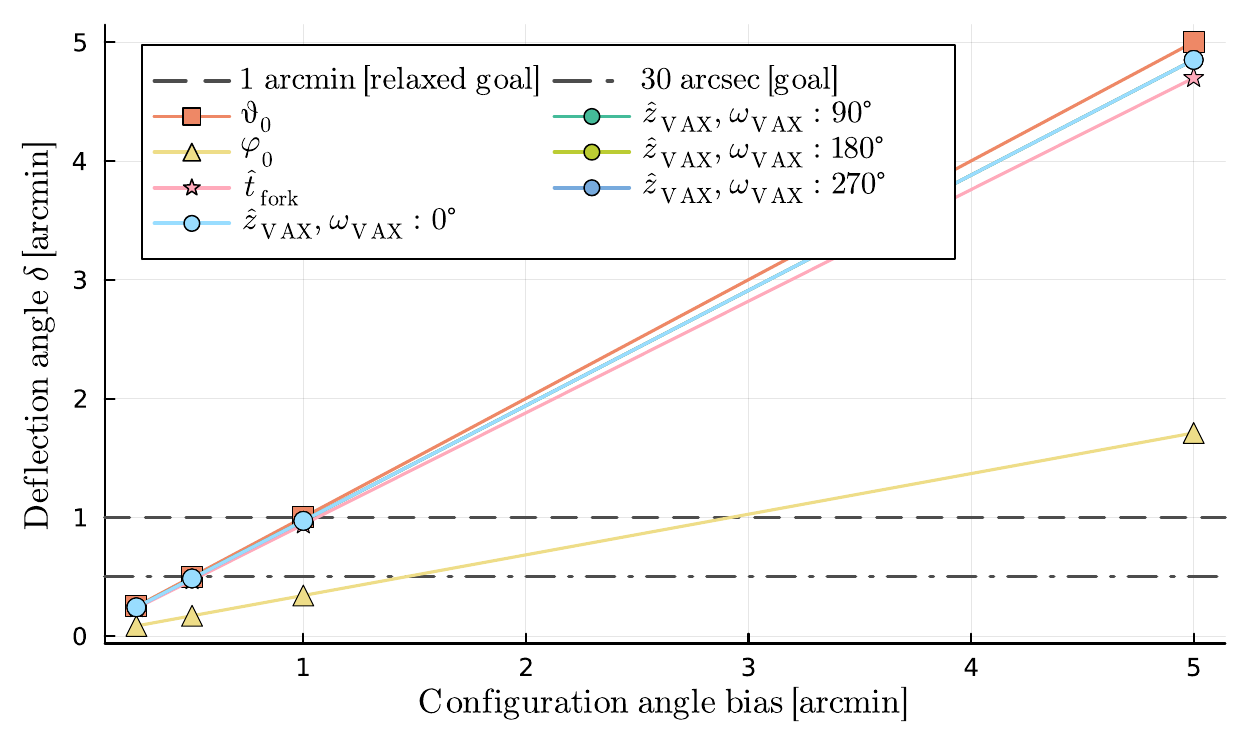}
    \caption{Pointing accuracy scaling. The figure shows the deflection angle averaged over 5~days as a function of the bias
    introduced in the configuration angles with values $15, 30, 60, 300~\arcsec$, four values of $\omegaVAX$ are used for $\zVAX\ne0\;\deg$ corresponding to the four cardinal directions. The horizontal dotted lines represent the required pointing accuracy in the nominal case ($30~\arcsec$) and a relaxed scenario ($1~\arcmin)$.  A  log-log scale is used to allow for proper data spacing.
}
    \label{fig:prm_errors_accuracy}
\end{figure}

The simulation of an entire survey for the whole instrument requires a considerable computational effort not justified by the scope of this study, which is to assess the overall effect of pointing error on the detected signal.
For this reason, after some initial tests, we decided to simulate blocks of 5~days of observation for the central horn for which $\panCam=\tiltCam=\rollCam=0\deg$ with a sampling frequency of $50~\Hz$.  
For each sample in the TOD, we computed the angular deflection defined in Eq.~\eqref{eq:deflP} between the biased and ideal pointings. 
We retrieved a linear relationship between each configuration angle and the resulting average deflection angle by running one simulation with a non-zero value per configuration angle. Our estimates are reported in Fig.~\ref{fig:prm_errors_accuracy}, which shows a linear scaling of deflection with an angular coefficient of nearly one for almost all configuration angles. The only notable exception is the azimuth offset $\varphiZero$, for which the deflection is smaller than the bias. We expect this behavior because, as specified in Eq.~(\ref{eq:deflP:small}), the deflection associated with an azimuth error is scaled by a factor $\sin\varthetaScan\approx 0.34$. Regarding the wobble angles, we see that the deflection only slightly depends on $\omegaVAX$ as the effect of $\omegaVAX$ uncertainty has to be scaled by $\sin \zVAX$.
Since the telescope will scan approximately one-third of the sky daily, five days of simulations are already representative of the whole scan. Indeed, we repeated part of the simulations at different epochs in the survey and noted that the results show relative variations of less than $10^{-3}$, which is an acceptable approximation for this work.

Afterward, we estimated the distortion of the expected signal from the sky due to the deflection. For this purpose, we used {\tt PySM~3} \cite{Zonca2021} to generate a Q/U Healpix map \cite{healpix_jl, Healpix} with $\Nside=512$; for simplicity, the sky map just included a power-law synchrotron spectrum at 43~GHz with isotropic spectral index. Using the ideal and biased pointing model, we used the map to simulate a scan using the ideal and the biased pointing model. For the latter, the value of each sample was retrieved along the \emph{biased} direction. Still, it was attributed by the map-maker to the pixel along the \emph{ideal} (unbiased) direction. Specifically, due to the finite resolution of the observed map, the value is derived by interpolating neighboring pixels using the \HEALpix\ {\tt interpolate} function. Therefore, the difference between a Q/U map generated for the case in which a bias is applied, and the Q/U map generated with zero bias provides a noise map for the pointing error and gives an estimate of the impact of pointing errors on the measurement of the sky signal. 

The noise maps in $\polQ$ and $\polU$ show some structure only on the Galactic plane, where the gradients of $\polQ$ and $\polU$ across a pixel are significant. It is possible to resume their properties by looking at the statistical distribution of noises. Given $\polQ$, $\polU$ error maps have very similar histograms,  instead of analyzing $\polQ/\polU$ maps separately, we converted them into a total polarization map $\polP = \sqrt{\polQ^2 + \polU^2}$ and used the error on $\polP$, $\errorP=\polPbiased - \polPideal$ as our figure of merit.

\begin{figure}[!h]
    \centering
    \includegraphics[width=\textwidth]{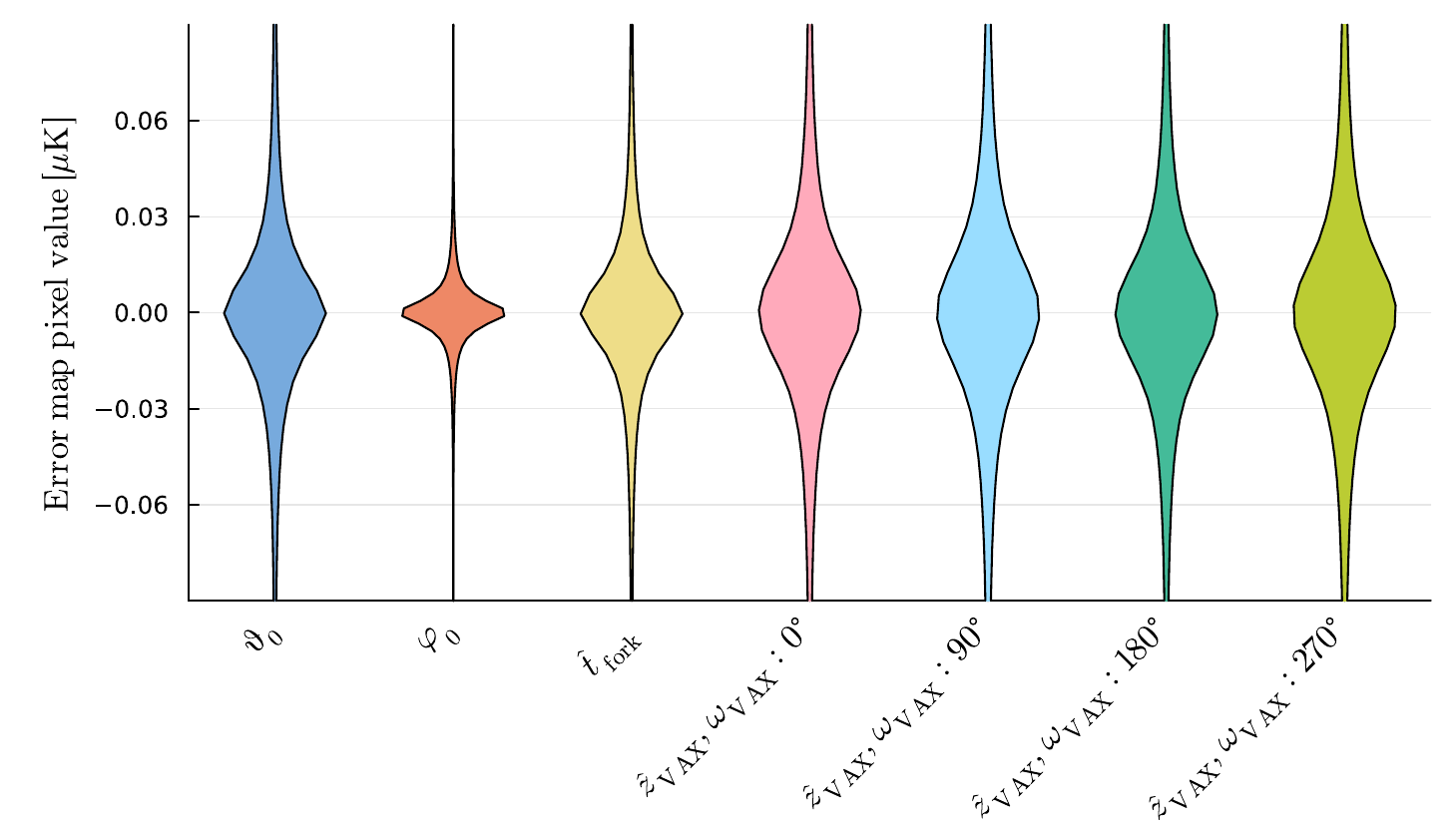}
    \caption{Error map pixel distributions. The figure shows the respective “violin plot” of the error distribution in $P$ across the map for every biased configuration angle. Configuration angles are biased by $1~\arcmin$, while as in  Fig.~\ref{fig:prm_errors_accuracy}, four values of $\omegaVAX$ are taken along the four cardinal directions.}
    \label{fig:prm_errors_histogram}
\end{figure}

\begin{figure}[!h]
    \centering
    \includegraphics[width=\textwidth]{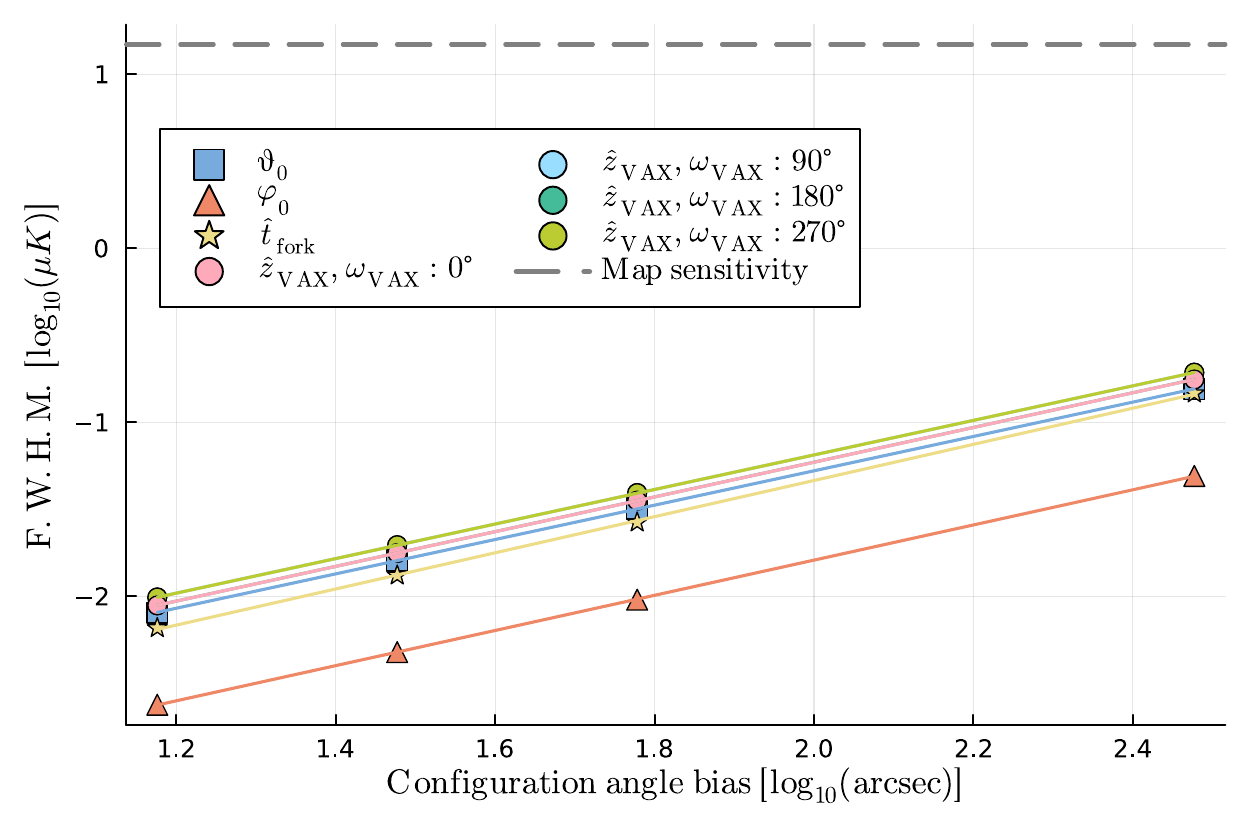}
    \caption{The figure shows how the F.W.H.M. of the distribution of errors in $P$ scales as a function of the bias. Tested biases are $[15, 30, 60, 300]~\arcsec$. The horizontal dotted line represents the level of white noise associated with the sensitivity of the instrument related to a sky map with $\Nside=512$.}
    \label{fig:prm_errors_width_scaling}
\end{figure}

Fig.~\ref{fig:prm_errors_histogram} shows the distributions of the $\errorP$ across the P maps for $1\;\arcmin$ bias for each varied configuration angle. 
Map averaged errors are below $10^{-5}\;\microK$ so that the distributions are zero-centered, symmetric with long tails, well described by a zero mean Cauchy distribution 
\begin{equation}
P(\errorPc) = \frac{1}{\pi} \frac{(\gamma_c/2)^2}{\errorPc^2 + (\gammaPc/2)^2};
\end{equation}
with $\gamma_c\approx 3\times 10^{-2} \; \microK$ the distribution Full Width at Half Maximum (FWHM) whose exact value depends on the configuration angle and the bias. 
To see how the $\gammaPc$ changes with the bias, we repeated the simulations for bias values equal to $15,\;30,\;60,\;300\;~\arcsec$ and got the results reported in Fig.~\ref{fig:prm_errors_width_scaling}. The figure (in log-log for graphical convenience) shows a good linear relationship between FWHM and the bias in the form:

\begin{equation}\label{eq:distribution:scaling}
    \gammaPc = \gammaPcOne 
    \left( \frac{\alpha_c}{60\;\arcsec}\right); 
\end{equation}

\noindent
where 
$\gammaPc$ is in $\microK$, $\gammaPcOne$
is the FWHM for $60\;\arcsec$ bias and 
$\alpha_c$ is the bias for the control angle $c$ in $\arcsec$. The values of
$\gammaPcOne$ for the different configuration angles 
are tabulated in Tab.~\ref{tab:coeff:bias}.
The configuration angle that creates the most significant error is $\zVAX$ followed by $\varthetaZero$ and $\tiltFork$. In contrast, the error introduced by $\varphiZero$ is dumped by the factor $\sin \varthetaScan$, as we have already pointed out in Fig.~\ref{fig:prm_errors_accuracy}, and $\omegaVAX$ has a small periodical effect. 
These results must be compared with the instrumental sensitivity rescaled to a map with $\Nside=512$: $\sigma_{512}=15~\microK$ \cite{thelspecollaboration2020large}, represented by the dotted line at the top of Fig.~\ref{fig:prm_errors_width_scaling}. The worst error for the most significant bias is less than $1.3\%$ of the pixel sensitivity. 
As expected from the geometry of the problem, the linear relationship between $\gammaPc$ and $\alpha_c$ breaks at large biases, but the level of non-linearity is below one percent, and it is negligible for our scopes.
At last, when independent biases on the configuration angles are considered, the resulting error has a Cauchy distribution with FWHM approximated by the sum in squares of the single $\gammaPc$. So, for a $1\;\arcmin$ bias applied to all the control angles $\gamma_P \lesssim 5.5 \times 10^{-2} \; \microK$.

\begin{table}[!h]
\centering
\begin{tabular}{lc}
Biased configuration angle $\alpha_c$         & $\gammaPcOne$ [$\microK$] \\ 
\hline
$\varthetaZero$                     & $2.81 \times 10^{-2}$\\
$\varphiZero$                       & $0.87 \times 10^{-2}$\\
$\tiltFork$                         & $2.65 \times 10^{-2}$\\
$\zVAX$ ($\omegaVAX=0^{\circ}$)     & $3.46 \times 10^{-2}$\\
$\zVAX$ ($\omegaVAX=90^{\circ}$)    & $3.76 \times 10^{-2}$\\
$\zVAX$ ($\omegaVAX=180^{\circ}$)   & $3.45 \times 10^{-2}$\\
$\zVAX$ ($\omegaVAX=270^{\circ}$)   & $3.76 \times 10^{-2}$\\ 
\end{tabular}
\caption{\label{tab:coeff:bias}
FWHM for $1\;arcmin$ bias in 
the configuration angles
for the simulations shown in Fig.~\ref{fig:prm_errors_width_scaling}.
Those are the coefficients of 
Eq.~(\ref{eq:distribution:scaling}).
}
\end{table}

In conclusion, the results show that the error associated with a bias of 1~arcmin in one of the configuration angles leads to a systematic error at least two orders of magnitudes smaller than the instrumental sensitivity. Thus, given the linear relation between biases and the pointing accuracy, we can state that the accuracy requirement could be relaxed up to 1~arcmin with a negligible impact on the observed signal.


\section{The Star Tracker}\label{sec:str}

The seven angles of the PRM could be entirely determined by observing sources in the sky whose angular diameter is much smaller than the beam $\FWHM$.
A point source with brightness temperature and solid angle $(\Tb, \OmegaSource)$ at the center of a beam of solid angle $\OmegaBeam$ produces a peak signal with an antenna temperature $\Tant = \OmegaSource/\OmegaBeam \Tb$ \cite{maris:etal:2021}. A source can be considered {\em point-like} if its angular size is at most some $10^{-2}$ of the beam size, leading to $\OmegaSource/\OmegaBeam< \mathrm{some} 10^{-3}$. Even a bright source as a planet, with $\Tb$ in the range of hundreds of $\Kelvin$, will result in signals of some hundreds of $\milliKelvin$ or smaller. 
It should also be considered that planets have very low levels of polarization at the frequencies of interest of \STRIPLSPE\ with upper limits below some fraction of a percent \cite{wmap:planets} equivalent to polarized antenna temperatures below one $\milliKelvin$. 
This has to be compared with the instrument's actual expected noise performance, the atmospheric noise, and the sky background.
Although the \STRIPLSPE\ radiometers are optimized to have very low $1/f$ noise in the polarization channel, the same is not true for total power observations, which are dominated by $1/f$ noise and whose performances are challenging to assess in this stage of instrument development. 
In short, we considered it unwise to base the entire PRM calibration on observing bright point sources. 
So we plan to mount a 
\textit{star camera} 
or 
\textit{star tracker} 
(STR) onboard \STRIPLSPE\ to provide the data needed to calibrate the PRM configuration angles by optical observations. If feasible, observations of bright sources like planets will be used as a further radio/optical intercalibration. Some quantitative results are given in Sect.~\ref{sec:prm:intercalibration}, while a more thoughtful analysis on the potential calibration of \STRIPLSPE\ by using sky sources is provided in \cite{Genova-Santos:etal:2023}.
 %
Given their importance in the space economy, the theory and practice of star trackers have been the subject of intense research; see, for example, \cite{Liebe:1993,Liebe:2002,howell:2006,Lang:etal:2010,Schwarz:2015,HMF:2019,IzGh:2019}.
In \STRIP\, the STR captures a frame of a sky field at some epoch of observation $\tobs$. It feeds a pointing reconstruction software that matches the frame with a catalog of stars to determine the direction it is aimed in the ICRF J2000 reference frame (the so-called {\em lost-in-space} problem). 
Using a standard astrometric library as the \astropy\ package, for a given epoch of observation $\tobs$ and the observer location, the $(\Alt, \Az)$ of the line of sight of the STR is determined from the reconstructed ICRF J2000 coordinates.
By repeating the above procedure for several combinations of control angles, the software compiles a matrix that links $(\vartheta, \varphi)$ to $(\Alt, \Az)$. This matrix can be used to determine the pointing model from the STR.
A complication arises because we cannot assume that the STR is precisely aligned with the telescope. Therefore, the PRM derived from the STR is correlated to the telescope PRM through two additional angles. The angles can be determined during a ground-based calibration campaign by observing a single bright radio point source, which can also be an artificial one, such as a UAV equipped with a radio beacon.

The main relevant requirements for the \STRIP\ relevant for the STR are:
i.)  the random error in pointing accuracy has to be of about 10~arcsec on a single frame, 
to ensure a negligible impact on the 1~arcmin error budget of the PRM calibration;
ii.) The field of view (FOV) must be at least $7\deg$, to ensure the coverage of a sufficiently large fraction of the focal plane of \STRIP;
iii.) limiting magnitude $\Vlim=7$~mag for an exposure time $\texp\approx0.01\;\sec$, as at least three stars have to be observed to solve the lost-in-space problem\footnote{By simulating STR observations at the telescope site we derived the following formula for the number of detected stars per frame; 
\begin{equation}
\left<\ndetectedstars\right> = 10^{0.505 \Vlim -3.1} \left( \frac{\FOV}{3\;\deg} \right)^{1.927},
\end{equation}
with $\Vlim$ the limiting magnitude. For $\Vlim = 7$ and  $\FOV = 7\;\deg$, we have $\sim$14 stars/frame, and 97\% of frames will have at least three stars.
} \citep{Liebe:1993,Liebe:2002,Lang:etal:2010,HMF:2019,IzGh:2019}; 
iii.) frame rate of at least 100 fps;
iv.) ability to synchronize the camera within $10^{-2}\;\sec$ with the UTC time.

The main parameters characterizing the camera are the quantum efficiency, $\quantumEff$, the sensor size $\LCAM$, and the pixel size $\pixelSize$. The objective is described by the diameter $\Diam$ and the focal length 
$\flength$.
The largest FOV of the system is $\fovAperture=2\arctan \LCAM/2/\flength$. The $\LCAM\ll \Diam $ is generally preferred to use the central part of the focal plane where distortions are minimized.
The accuracy in determining a single star position is proportional to the S/N for the detected star which is proportional to $\sqrt{\quantumEff \Aopt F_{*} \texp}$, where $\Aopt\propto\Diam^2$ the optics effective collecting area, $F_{*}$ is the number of photons from the star hitting the detector per unit time and $\texp$ the exposure time \cite{howell:2006}. A S/N>4 is required to have a good position determination.
Given $\texp \in 1 \div 10^{3} \; \msec$ the accuracy is determined $\quantumEff$ and $\Diam$ \cite{Liebe:2002}. 

If just one star were observed and the optics were able to produce a point-like image, the accuracy of the star tracker would be fixed by the size of the pixel projected in the sky $\pixelSizeSky = \pixelSize/\flength$. 
Assuming the star is projected randomly within the pixel, the RMS of the pointing error will be $0.38\pixelSizeSky$ and given $\ndetectedstars>3$, the pointing error will be reduced by a factor $1/\sqrt{\ndetectedstars}$ \cite{Liebe:2002}.  %
Diffraction, seeing, and other atmospheric or optical defects widen the star image, and in practice, the attainable accuracy quoted in literature is in the limit $0.05-0.2\;\mathrm{pixels}$ \cite{Liebe:2002}.
Therefore, a typical configuration with a pixel size $\pixelSize=5\;\micron$ coupled with a $\flength=100$~mm objective leads to a theoretical pointing accuracy $\approx2 \div 5\;\arcsec$.


\subsection{Implementation}\label{sec:str:implementation}

Using the design criteria quoted above, a prototype for the \STRIPLSPE\ STR was implemented starting from components available in 2018. 
The accepted baseline configuration comprises a Samyang Lens 85mm-F1.2 and a ZWO ASI 174MMC camera.
The design pixel scale for this coupling is $14.2\;\arcsec/pxl$, and the f.o.v. is $7.64\deg \times 4.80\deg$.
The expected S/N for a $V=7$ star and $\texp=0.01\;\sec$ is 4; with this limit, an average of 14 stars/frame are expected to be detectable.
The expected pointing accuracy in ideal conditions is about $2 \div 5\;\arcsec$.

\begin{figure}
\centering
\includegraphics[width=\textwidth]{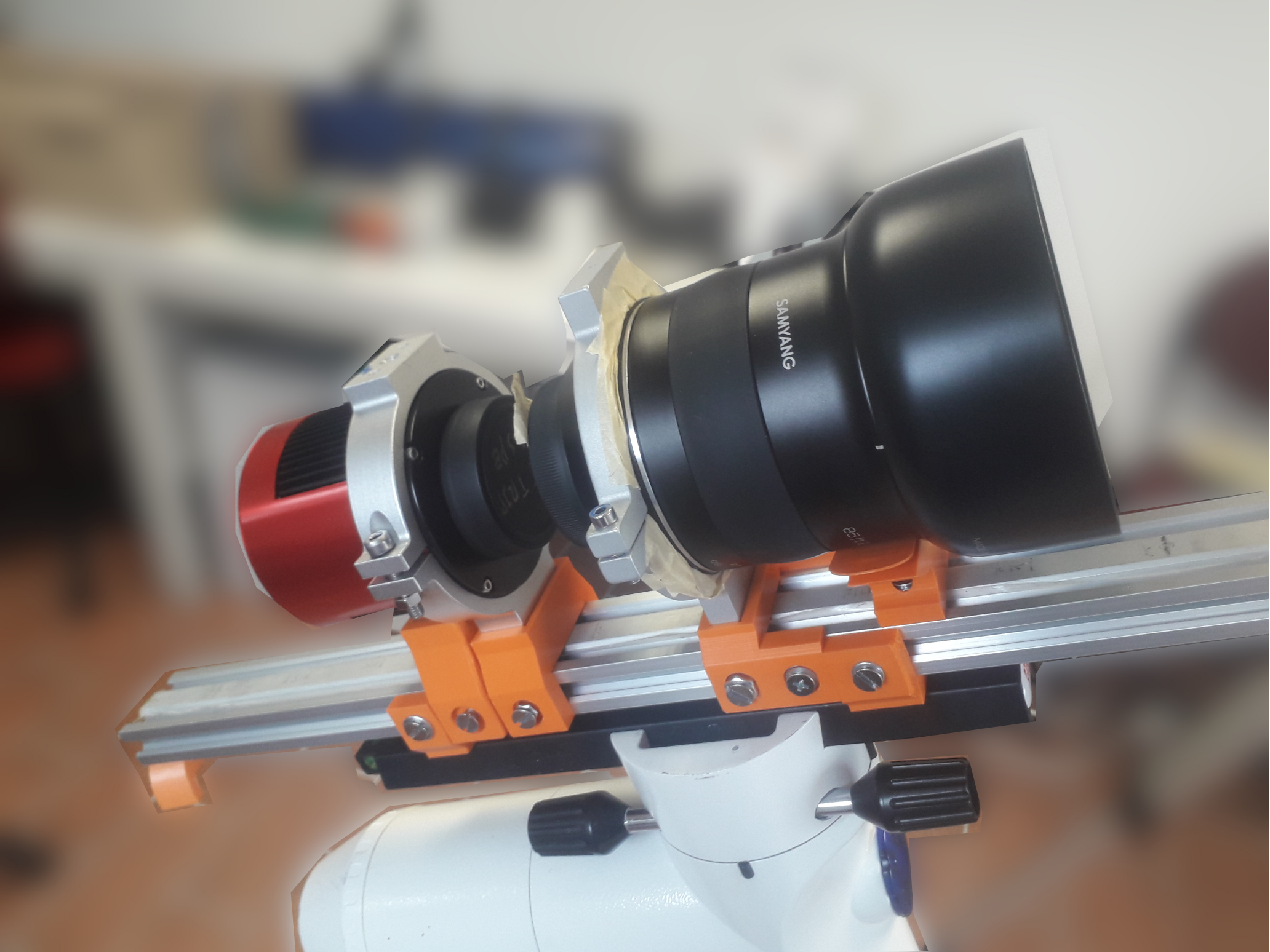}
\caption{\label{fig:str:proto}
The STR prototype used for the test campaign is described in this paper.
}
\end{figure}

\begin{table}
\centering
\begin{tabular}{ll|ll}
\multicolumn{2}{c}{Camera}  & \multicolumn{2}{c}{Objective} \\
\multicolumn{2}{c}{ZWO ASI 174 MMC }        & \multicolumn{2}{c}{Samyang XP 85mm F1.2} \\
Producer          & ZWO                     & Samyang           &                      \\
Technology        & CMOS                    & f                 & 85 mm                \\
Sensor            & SONY IMX174LLJ          & D$^{(ii)}$        & 70 mm                \\
Pixel Size        & $5.86\;\micron$         & f/D               & 1.2                  \\
Sensor Format     & $1936\times1216 \;\pxl$ & Diaphragm control & electronic           \\
Sensor Size       & $11.34\;\mm$            & Focus control     & Manual               \\
RON               & $3.5\;e^{-}$            & Mount             & Canon EF             \\
QE                & $78\%$                  & Body diameter     & 93 mm                \\
FPS max           & $164\%$                 & Body weight       & 1050 gr              \\
Shutter           & Global                  &                   &                      \\
Exposure Range    & $32\;\mu\sec$-1000$\sec$&                   &                      \\
Data Interface    & USB3.0                  &                   &                      \\
GPIO              & None                    &                   &                      \\
Power supply      & 12 VCC, 2 A max         &                   &                      \\
Cooling           & Regulated two-stage TEC &                   &                      \\
Cooling $\Delta T^{(i)}$& $35\Celsius$ -- $40\Celsius$&         &                      \\
ADC               & 8 and 10 bits           &                   &                      \\
Weight            & 450 gr                  &                   &                      \\
\end{tabular}
\caption{\label{tab:selected:configuration}
Selected configuration for the prototype STR.
Notes: i) $\Delta T$ is the most significant temperature difference reachable between the sensor and the environment;
ii) $D$ the aperture is inferred from the focal length and the focal ratio.
}
\end{table}

\begin{figure}
\centering
\includegraphics[width=\textwidth]{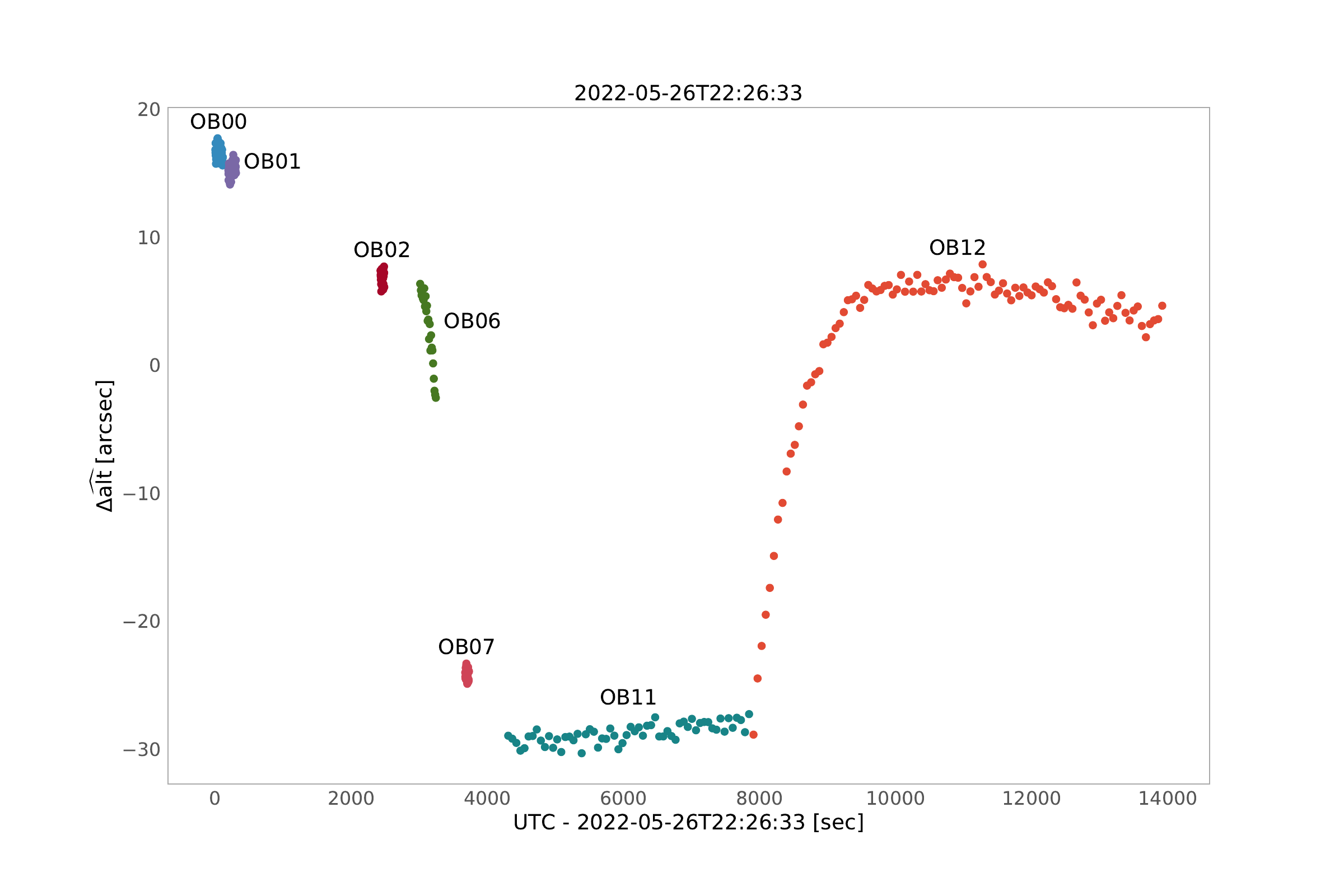}
\caption{\label{fig:temp:jump}
A cooling test was carried out on the night of 2022-05-26. The plot shows the change in pointing elevation for the STR aimed in a fixed direction as a function of time.
The $\Delta \Alt$ is computed with respect to the night mean of $\Alt$.
The different colors refer to blocks of observations with varying temperatures of the camera:  
OB00, OB01 (blue and violet) cooling off, camera at $\approx+16\Celsius$; OB02 camera (brown) cooled at $\approx+10\Celsius$; OB06 (dark green) temperature drifting from $+10\Celsius$ down to $\approx0\Celsius$; OB07 (light red) camera at $-30\Celsius$; OB08 (light green) programmed temperature $\approx-40\Celsius$, the factory limit for the given ambient temperature was $-35\Celsius$; OB12 (light red) temperature control off, temperature drifted from $\approx-40\Celsius$ to $\approx+10\Celsius$.
} 
\end{figure}

Fig.~\ref{fig:str:proto} shows the prototype assembly. A {\tt Canon EF-T Mount} adapter joins the camera and the optics.
Two aluminum half rings and some PLA supports join the camera. The objective is an aluminum bar that provides the needed stiffness to the system. This bar is connected to a {\tt Vixen} bar, which allows to host the camera on a {\tt Vixen Sphynx} equatorial mount. The aluminum bar is also used to support other components, such as a temperature probe, the cables, and the time synchronization pulsating source (not shown in the image).

The rest of the front end (i.e., the support electronics, the power supply, the control DPU, and the synchronization and environmental MPUs) are housed in a separate IP54 box.
The control DPU is an {\tt NVIDIA Jetson Nano Developer Kit}
board\footnote{\url{https://developer.nvidia.com/embedded/jetson-nano-developer-kit}}.
A custom version of the \Linux\ operating system operates the board.
The camera is connected to the DPU through the USB3 port 1.
The time synchronization MPU is based on an {\tt Arduino ZERO}\footnote{\url{https://store.arduino.cc/collections/boards/products/arduino-zero}}
module and a u-blox GPS module that are commonly used onboard UAVs.
The Arduino ZERO is connected to the DPU through the USB3 port 2.
The GPS delivers a Pulse per second PPS with an accuracy of 10\,nsec to the \Arduino\ GPIO and NMEA packets
through a TTY serial connection.
The ENV MPU is based on an
{\tt Arduino MKR Zero}\footnote{\url{https://docs.arduino.cc/hardware/mkr-zero}}
connected to the DPU through the USB3 port 3. This prototype used the ENV MPU to read temperature and humidity sensors and monitor dangerous environmental conditions.
The Camera Control Software executed by the DPU is written in \PythonTh\ with a \Cpp\ kernel based on the
{\tt ASI Linux Mac SDK V1.16.3}\footnote{https://astronomy-imaging-camera.com/software-drivers} patched with the Open-Source
{\tt ASI Camera Boost} developed by P. Soya\footnote{\url{https://github.com/pawel-soja/AsiCamera}} to
amend the low frame rate allowed by the original library.

\subsection{Synchronization}\label{sec:str:sync}

A drawback of the selected camera is the lack of a GPIO with a pin signaling the start and end of a frame acquisition. This pin could be used to provide precise timing for each frame. However, affordable cameras on the market could not simultaneously ensure GPIO and thermal stabilization. We prefer to buy a camera that can be thermally stabilized, providing ourselves with a time synchronization method.

The importance of thermal stabilization is shown by the extreme case presented in Fig.~\ref{fig:temp:jump}.
The test procedure was similar to the one illustrated in Sect.~\ref{sec:str:testing}: 
the camera was pointing in a fixed direction and
different camera configurations were tested, each one being labeled in the plot.
In OB00 and OB01, the cooling system was left off, while in OB02, it was turned on, fixing the sensor temperature at $\approx10\Celsius$, i.e., $6\Celsius$  below ambient
temperature. In this test, the frame acquisition started when the camera reached the required temperature. In OB06, another decrease of $\approx10\Celsius$ was programmed,
but frames were acquired during cooling. In OB07, the camera was instructed to decrease the temperature to $\approx-30\Celsius$, near the limit of the maximum allowed
cooling with respect to the ambient temperature. In OB11, the camera was programmed to the maximum possible temperature decrease. In this state, the camera did not have
optimal temperature stabilization, and the frame center drifted slightly as the sensor temperature followed the ambient temperature. 
At last, in OB12, the cooling system was turned off, leaving the sensor to warm up. In conclusion, these data show that we have approximately $1.25\;\arcsec/\Celsius$ of pointing drift for a change
in the temperature sensor. 

\begin{figure}
\centering
\includegraphics[width=0.9\textwidth]{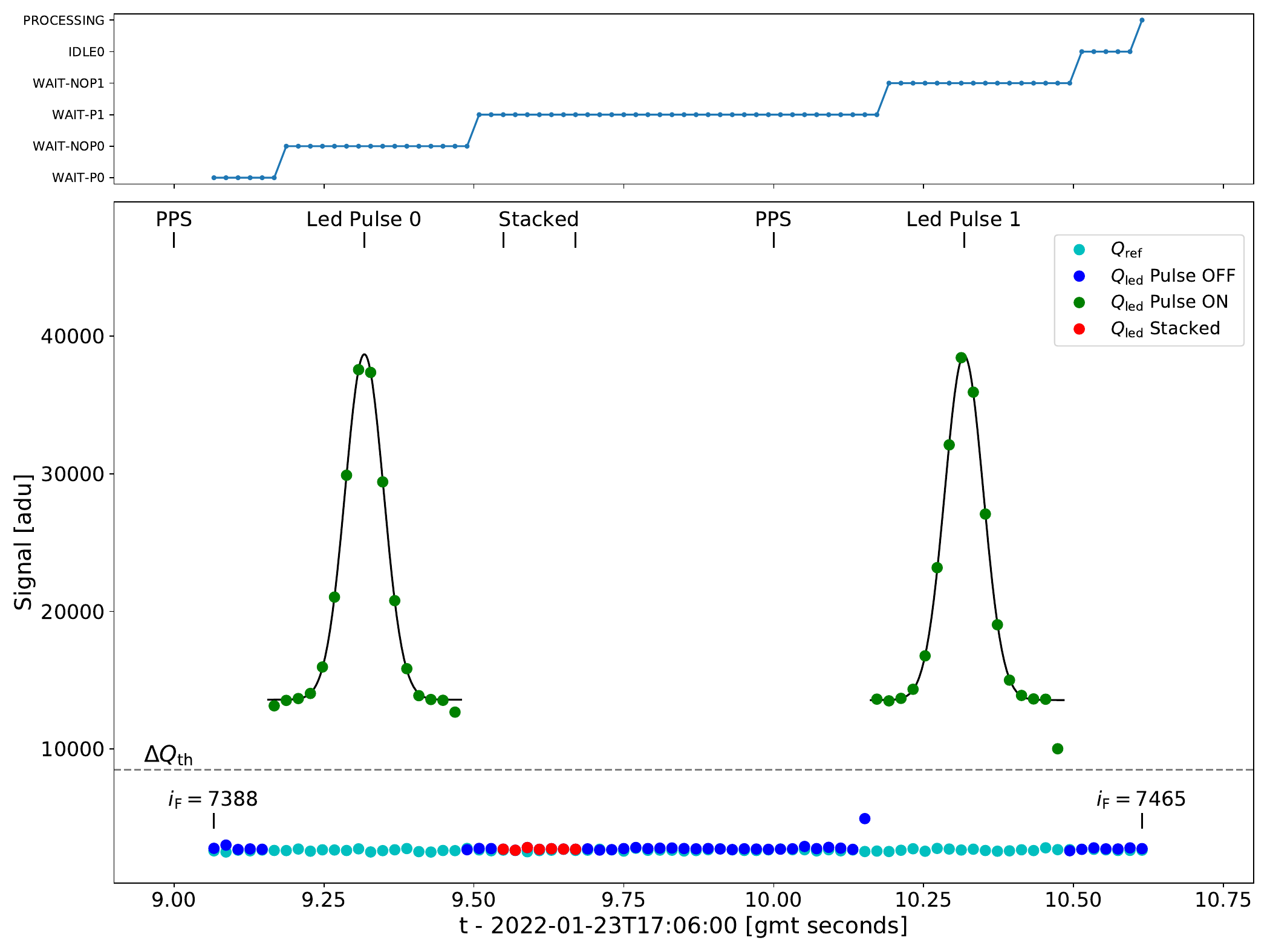}
\caption{\label{fig:pps:sync}
Frame Synchronization Scheme.
The figure refers to the synchronization of an observation taken at GMT 17:06:09 of 2022-01-23.
The two plots share the same horizontal scale, which is already calibrated in seconds after GMT 17:06:00.
The upper plot represents the sequence followed by the state machine model of the camera software.
The lower plot represents the synchronization. In the lower plot
cyan circles are measures of $\refSignal$,
green circles represent
$\ledSignal$
for frames where the camera software has detected the pulse, red circles are for samples used to create the stacked image, and blue circles are for the other frames.
The gray line is the fitted Gaussian model plus the pedestal for each pulse.
At the top of the figure, markers identify the two PPS, the center of each pulse, and the time interval spanned by stacked frames.
The frame numbers for the first and last frames are at the bottom. The gray dashed line is the threshold value for discriminating the LED signal.
}
\end{figure}

To allow a camera synchronization with UTC 
within $10^{-2}\;\sec$,
we implemented an external synchronization mechanism by sending light pulses to the camera
which are detected by the Camera Acquisition Software.  
The camera acquires frames in {\em free-run} mode at a constant rate, each identified by a frame number $\frameIndex$.
Frames are then synchronized off-line 
by compiling a table of UTC times of pulses, $\tpulse$, and the corresponding $\frameIndex$ of the frames in which they are seen. 

The parameters of the scheme are  
$\texp$, the interval between two consecutive frames $\frameStep$  and
the with of a pulse $\pulseWidth$.
Of course  
if $\pulseWidth < \frameStep$ or even worst $\pulseWidth < \texp$ some pulses will be not observed, while 
if $\pulseWidth > \frameStep$ the ultimate accuracy in synchronization will be fixed by $\pulseWidth$.
 To overcome this problem, we provided a pulse with a well-defined and measurable shape,
so that the pulse seen in the frame has the form 

\begin{equation}
\cameraV(\pulseShape) = \cameraGain \pulseShape(t)+\cameraOffset;
\end{equation}

\noindent
with $\pulseShape(t)$ the pulse shape.
By fitting the shape of the corresponding light curve $\cameraV(\pulseShape(t))$, it will be possible to determine the position of the center of the pulse using $\frameIndex$ as a proxy for time. From $\tpulse$, it will be possible to perform the time calibration.

The implementation of this scheme needs accurate pulse generation and formatting.
The pulse was generated by a 5\,mm red LED placed in front of the lens, illuminating only a portion of the frame. The SYNC MPU drives the LED. 
Given the high sensitivity of the camera, the LED must produce a minimal amount of light when it is fully ON so as not to saturate the sensor.
Typical LED currents are of the order of some tens to some hundreds of $\microAmpere$, which the output pin of the MPU can easily manage.

To shape the brightness of the pulsed signal, we exploited the Pulsed Width Modulation technique (PWM), which can be straightforwardly implemented in an MPU \cite{arduino:cookbook}, \cite{avr:programming}.
The PWM exploits that the sensor integrates the LED source for a finite time $\texp$.
The LED is driven by a square wave of period $\pulseStepHF \ll \texp$, which turns the LED on for a fraction of the period
$\pulseStepHFDuty$, named \emph{duty cycle}.
The amount of light collected is then proportional to: $\pulseBrigthOn\pulseStepHFDuty\texp$ with $\pulseBrigthOn$ the LED brightness,
so modulating $\pulseStepHFDuty$ it is possible to modulate the amount of light the camera receives.
The wanted $\pulseShape(t)$ is tabulated in discrete steps of $\pulseStepHFDuty$ with duration $\pulseStepLF$.
Of course we must have $\pulseStepHF \ll \pulseStepLF<\texp$, having typical $\texp>10^{-2}\;\sec$, $\pulseStepLF = 10^{-3}\;\sec$ and
$\pulseStepHFDuty = 10^{-5}\;\sec$.
 The MPU uses its internal clock $\tMPU$ as a time base for the generation of pulses; $ \tMPU$ is simply the number of DPU ticks that have elapsed since the start of the program.
 The cadence of a DPU tick is a multiple of the actual DPU clock cadence; in our case, we selected a DPU tick cadence of approximately $2\microsec$\footnote{The \Arduino\ used in this case has a 48~MHz clock, which is scaled by 96.}.
 The MPU is connected to an external precise clock sending a pulse-per-second (PPS) at each second of UT time.
During our tests, the external clock was represented by a u-blox GPS module whose expected accuracy in generating the PPS is approximately 10\,ns, which simulates the internal telescope master clock.
Together with the PPS, the MPU receives an NMEA packet needed to identify the UTC of the second with which the pulse is associated.
The MPU stores the time of its internal clock at which each PPS is received $\tMPUPPS$.
The train of PPS received by the MPU split the time interval into time frames, each identified by the UT time of the starting pulse.
Within each time frame, the MPU is programmed to generate a single short pulse
which is the discretized version of a constant pedestal plus a Gaussian pulse.
To allow the MPU to handle the PPS and the associated NEMA packet properly, a fixed delay
$\delayPulseTrigger\approx0.1\;\sec$ is applied between the reception of the PPS and the LED pulse generation.
Denoting with $\tilde{t}=\tMPU-\tMPUPPS$ the MPU time elapsed from the last PPS,
After receiving the PPS, the LED brightness is modulated with the law

\begin{equation}\label{eq:pulse:shape}
\pulseShape(\tilde{t}) =
\left\{
\begin{array}{ll}
 0 \le \tilde{t} < \delayPulseTrigger; \\
 \pulsePedestal + \pulsePeak \exp(-(\tilde{t}-\pulsetPeak)^2/2\pulseSigma^2), \delayPulseTrigger \le \tilde{t} \le \delayPulseTrigger+\pulseWidth; \\
 0, \delayPulseTrigger+\pulseWidth < \tilde{t};
\end{array}
\right.
\end{equation}

\noindent
where $\pulseWidth$ is the time interval over which the LED is ON,
$\pulsePedestal$ and $\pulsePeak$ the LED brightness pedestal and the LED peak brightness, $\pulsetPeak$ the position of the peak in the time-frame and the $\pulseSigma$ the shape parameter of the Gaussian pulse. The sequence ends with the LED off until the next PPS is detected.
In all our tests, $\pulsetPeak=\delayPulseTrigger+\pulseWidth/2$ but other choices can be considered.
For each time frame, the MPU sent in output $\tMPUPPS$ of the first and the second pulse together with the corresponding NMEA UTC times. In addition, the $\tMPU$ corresponding to the beginning of the pulse, the peak, and the end of the pulse are also sent.

The Camera Acquisition Software
detects the synchronization pulse and generates \FITS\ files with the astrometric images.
There is no need to store frames at a rate of 100 fps so that 
the Camera Acquisition Software outputs $\naver$ averaged stacked frames, which are saved as \FITS\ files.
Of course, we discard those frames where we detect a pulse, but the Camera Acquisition Software
saved the observed LED brightness in a separate file for each acquired frame.
This file is used to synchronize the averaged frame during the offline analysis.

Fig.~\ref{fig:pps:sync} shows an outline of the process for an observation taken on 2022-01-23 at 17:06:09 GMT.
The upper plot of Fig.~\ref{fig:pps:sync} represents
the Camera Acquisition Software as a state machine traversing the sequence of states:
\statusWAITPZERO,
\statusWAITNPZERO,
\statusWAITPONE,
\statusWAITNPONE,
\statusIDLEZERO,
\statusPROCESSING.
The CAS stores the frames in a circular register, assigning them the corresponding frame index $\frameIndex$.
In the first state, \statusWAITPZERO, the CAS is looking for the pulse. If not detected, the frame in the circular buffer is tagged as
{\em pulse-off}.
If a pulse is detected, the frame is tagged as {\em pulse-on}, and the CAS
 enters the \statusWAITNPZERO\ status.
The CAS remains in this state until a frame without a pulse is detected. At this point, the frames are tagged as {\em pulse-off}, and the camera enters the
state \statusWAITPONE,
until the LED switches again to on. 
When this happens, the CAS enters
\statusWAITNPONE\ remaining there until the second pulse is turned off.
At this point, the CAS enters
\statusIDLEZERO, lasting few frames, after which switches to 
\statusPROCESSING. When processing is concluded,
the CAS idles for less than ten frames and then backs to \statusWAITPZERO.

In \statusPROCESSING\
state $\naver$ consecutive frames taken in \statusWAITPONE\ are averaged and saved in a \FITS\ file, together with the time sync information.
In general, the number of frames acquired in the \statusWAITPONE\ state is larger than
$\naver$, so more averaged images can be generated. As most of the time needed for processing is spent in writing the \FITS\ on the mass memory,
during the test
in Fig.~\ref{fig:pps:sync}
the software has been instructed to produce only one stacked frame.
The \FITS\ header contains the frame index of the first and the last frame in the average.
The sync information is a table containing information for some frames acquired before the acquisition of the first pulse to some frames after the end of the second pulse. Each line of the table corresponds to a frame with the following information: the frame index, the status of the CAS, the strength of the pulse signal in the frame, the incremental pulse number from the beginning of the acquisition
and whether or not the frame has been used to generate the \FITS.

The images saved in \FITS\ files are the sum of $\naver$ frames so that the total exposure time is
\begin{equation}
\texptotal = \naver \texp;
\end{equation}
the frame in Fig.~\ref{fig:pps:sync} has $\texp=0.02\;\sec$ and $\naver=7$, resulting in a $\texptotal=0.14\;\sec$. This is a special case, as usually $\texp=0.01\;\sec$, $\naver=10$ and $\texptotal=0.1\;\sec$.

The signal in the lower part of Fig.~\ref{fig:pps:sync} is the pulse strength of a frame defined as
\begin{equation}\label{eq:pulse:strength}
\pulseStrength = \ledSignal-\refSignal,
\end{equation}
where $\ledSignal$ and $\refSignal$ are the average brightness measured within two regions-of-interest (ROI) of the image: the led ROI, which is drawn into the portion of the image illuminated by the LED, and the reference ROI, which is not illuminated by the led. Both $\ledSignal$ and $\refSignal$ are averages computed on the raw frames provided by the camera, so they are expressed in $\ADU$.
 To reduce the effect of noise, we apply a threshold $\thSignal$ so that the condition to tag a frame as one with {\em pulse detected} is
\begin{equation}\label{eq:pulse:detection}
\thSignal < \pulseStrength.
\end{equation}

The lower plot of Fig.~\ref{fig:pps:sync} shows for each frame the measured
$\refSignal$ values as cyan circles and $\ledSignal$ values as either blue circles for frames tagged as having the led-off and green for frames tagged as led-on.
The red circles are the frames used to generate a stacked frame saved as a \FITS\ file.
On the top of the figure, the position of the two PPS and the corresponding centers of the pulses as provided by the \Arduino\
ticks are shown.
The gray line overlapped with the green points represents the best-fit model for each pulse.

\begin{table}[t]
\centering
\begin{tabular}{lcrrrr}
              & \multicolumn{1}{c}{Pulse 0} & \multicolumn{1}{c}{Pulse 1} & difference  & unit\\
\hline
$\tpps$       & 1642957569                  & 1642957570                  & 1           & gmt seconds\\
$\tickspps$   & 2791339956                  & 2791840029                  & 500073      & ticks \\
$\pulsetPeak$ & 2791498663                  & 2791998738                  & 500075      & ticks \\
$\tpls$       & 1642957569.3173676          & 1642957570.3173714          & 1.0000038   & sec gmt \\
$\iFpls$      & $7400.480 \pm 0.004$        & $7450.231 \pm 0.004$        & $49.752 \pm 0.006$ & frames\\
\end{tabular}
\caption{\label{tab:pps:sync}
Synchronization for Fig.~\ref{fig:pps:sync}.
The table gives for each pulse:
   the GMT time $\tpps$,
   the MPU clock time when the PPS was received $\tickspps$,
   the MPU clock time for the center of the pulse $\pulsetPeak$,
   the GMT time of each pulse peak $\tpls$,
   the center of each measured peak $\iFpls$ with its uncertainty.
}
\end{table}

The relevant parameters for time synchronization are reported in Tab.~\ref{tab:pps:sync}.
There, the GMT $\tpps$ of each peak is expressed in seconds as UNIX time.
The MPU times are expressed in MPU ticks as provided by the MPU.
$\tickspps$ is the clock ticks of each PPS,
and $\pulsetPeak$ is the clock ticks of the center of each led pulse.
From the couples $(\tppsZ,\ticksppsZ)$ and $(\tppsO,\ticksppsO)$, the MPU time scale conversion to GMT is straightforward:

\begin{equation}\label{eq:mpu:sync}
t(\tilde{t}) = \frac{\tppsO-\tppsZ}{\ticksppsO-\ticksppsZ}(\tilde{t}-\ticksppsZ)+\tppsZ
\end{equation}

\noindent
this is used to compute the GMT time of the center of the two pulses $\tplsO$ and $\tplsZ$ from the corresponding
$\pulsetPeakZ$ and $\pulsetPeakO$.
Ideally, the difference between $\ticksppsZ$ and $\ticksppsO$ would have to be equal to $\pulsetPeakZ$ and $\pulsetPeakO$, but according to
Tab.~\ref{tab:pps:sync} they differ by two ticks, equivalent to $\approx 4\;\microsec$, which represents a negligible uncertainty.
From the two couples $(\iFplsZ,\tplsZ)$ and $(\iFplsO,\tplsO)$ the conversion from frame index $\frameIndex$ to GMT time is straightforward too having:

\begin{equation}\label{eq:frame:sync}
t(\frameIndex) = \frac{\tplsO-\tplsZ}{\iFplsO-\iFplsZ}(\frameIndex-\iFplsZ)+\tplsZ.
\end{equation}

\noindent
with uncertainty

\begin{equation}\label{eq:frame:sync:unc}
\sigma_t(\frameIndex) = \frac{\tplsO-\tplsZ}{(\iFplsO-\iFplsZ)^2}
   \left(\left|\frameIndex-\iFplsO\right| \erriFplsZ
   +
   \left|\frameIndex-\iFplsZ\right|\right) \erriFplsO
\end{equation}

\noindent
Having $\erriFplsZ\approx\erriFplsO\approx\erriFpls$ the error is  upper bounded by

\begin{equation}\label{eq:frame:sync:unc:upper}
\epsilon_t < \frac{\tplsO-\tplsZ}{\iFplsO-\iFplsZ}\erriFpls;
\end{equation}

\noindent
which for our case gives $\epsilon_t < 10^{-4}\;\sec$.


\subsection{Testing the prototype STR}\label{sec:str:testing}

The prototype version of the STR has been assembled and tested at the INAF/Trieste Astronomical Observatory in SW and HW through 2021 and 2022, with some aid provided by external laboratories.
During this period, we conducted a test campaign covering various functional aspects, mainly the Camera Acquisition Software and the data reduction.
To minimize mechanical perturbations to the camera caused by continuous mounting and dismounting of its parts, the prototype remained outdoors, exposed to the winter conditions for most of the time, apart from periods in which it was dismounted to implement some improvement or to protect it from terrible weather. The 2021/22 winter has been arid, with scarce wind for most of January\footnote{Trieste is known to have frequent episodes of {\em Bora} wind. A katabatic wind from the east, whose peak velocities frequently exceed 100 Km/h.}.

Functional tests were conducted during nightly observation runs.
After a setup phase, the camera was left unattended to execute a preprogrammed sequence of observations. The quality of the acquired images and the weather conditions were monitored through a remote connection.
At the end of the night, the acquired data were downloaded from the OBC storage, and after a simple sanity check on the frames, the data analysis pipeline was started.

The pointing accuracy and stability were determined in a dedicated functional test conducted over the nights from January 21st to January 30th.
For this test the Star Tracker was located on the flat roof of the building housing the {\em Reinfelder} telescope of INAF/Trieste Astronomical Observatory, in the town of Trieste, Italy \cite{Abrami:Cester:1959}\footnote{\url{https://www.oats.inaf.it/index.php/en/177-observatory/426-historical-notes-basevi.html}}.
The geographical coordinates of the camera were obtained with an accuracy of about 3\,m from the GPS used for time synchronization. They are $\Lat=45.644036\pm3\times10^{-5}\;\deg$~North, $\Long=13.774087\pm3\times10^{-5}\;\deg$~East, $h=73\pm3\;\meter$.
The star tracker was aimed at a fixed position in the sky toward the west, approximately at $\Alt=45\deg$; the sensor was rotated to have its long side roughly normal to the horizons. In these conditions, stars crossed the frame approximately diagonally.

The results of each test night are summarized in Tab.~\ref{tab:nights:observations}, giving the epoch of each observing night, the average pointing direction, and the estimated pointing accuracy. The night of January 27th was discarded due to bad weather conditions.

\begin{figure}
\centering
\includegraphics[width=\textwidth]{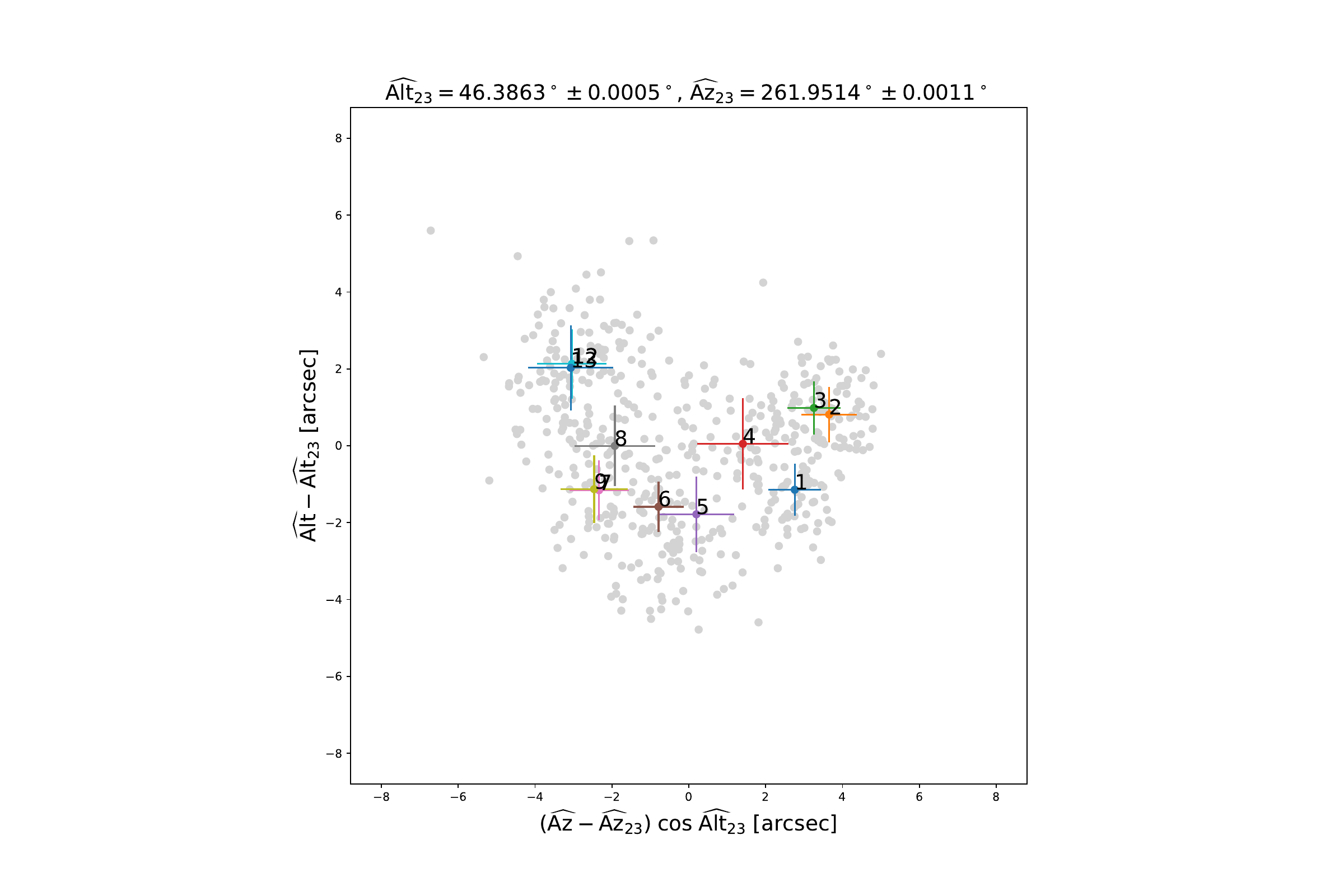}
\caption{\label{fig:night:alt:az:scatter}
Scatter plot of alt-az coordinates for the observations of 2022 Jan 23rd.
Numbers refer to different OB as listed in Tab.~\ref{tab:night:observing:blocks}; the error bars are the OB standard deviations.
OBs 10 and 11 have been discarded for too poor observing conditions or less than five useful frames.
}
\end{figure}

An observation run was divided into observing blocks (OB). The camera was left free to operate in free-run mode in each OB with a fixed setup. 
An example of observing rung is in Tab.~\ref{fig:night:alt:az:scatter}. 
The Camera Control Software resets the Camera when a new setup is provided, turning off the cooling system. Hence, each OB starts turning on the cooler and stabilizing the sensor temperature. This means that the first minute of each OB cannot be used.  Moreover,  frames acquired with a camera temperature of 0.1-0.2~K different from the target temperature were discarded in the subsequent data analysis.
Due to the limited storage space in the prototype OBC, during the test, stacked frames were produced at a cadence of one frame per minute, apart from OBs \#2 and \#3, for which a $10\;\sec$ cadence was used. 

\begin{table}
\centering
\begin{tabular}{cccccc}
Night & \#Fr & $\Alt_n$ & $\deltaAlt_n$ & $\Az_n$ & $\deltaAz_n$ \\
      &       & [deg]   & [arcsec]     & [deg]  & [arcsec]    \\
\hline
2022-01-21/22 & 449 &  46.38823 & 3.91 & 261.94929 & 1.97\\
2022-01-22/23 & 277 &  46.38429 & 2.23 & 261.94982 & 2.80\\
2022-01-23/24 & 514 &  46.38628 & 1.97 & 261.95138 & 3.83\\
2022-01-24/25 & 387 &  46.38474 & 3.70 & 261.94947 & 1.97\\
2022-01-25/26 & 137 &  46.38496 & 3.14 & 261.95062 & 3.78\\
2022-01-26/27 & 127 &  46.38498 & 1.47 & 261.95088 & 2.46\\
2022-01-28/29 & 360 &  46.38775 & 2.35 & 261.94777 & 1.83\\
2022-01-29/30 & 340 &  46.38569 & 1.69 & 261.94641 & 3.12\\
2022-01-30/31 & 494 &  46.38521 & 1.99 & 261.94630 & 3.91\\
\end{tabular}
\caption{\label{tab:nights:observations}
Table of observations for the January 21st - 31st test.
\#Fr\ is the number of accepted frames,
$\Alt_n$, $\deltaAlt_n$ , $\Az_n$ and $\deltaAz_n$ 
are the averaged $\Alt$, $\Az$ and their standard deviations for night $n=21, 22, 23, \dots, 30$.
}
\end{table}

\begin{figure}
\centering
\includegraphics[width=\textwidth]{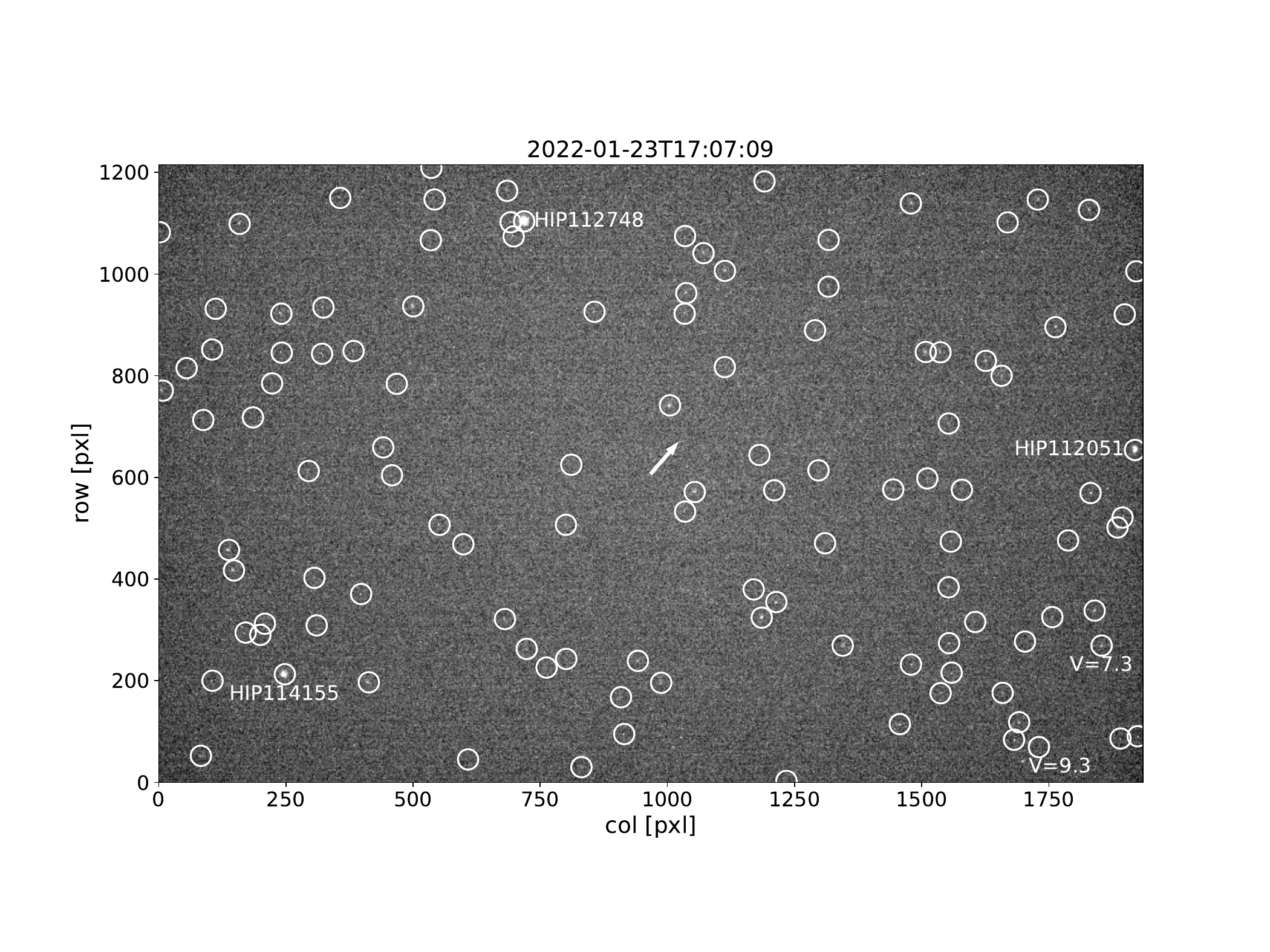}
\caption{\label{fig:night:frame:id}
Identification of stars for one frame of Night 2022 Jan 23rd.
The 114 circles mark detected objects that are identified in the Tycho2 catalog. The three brightest stars: 
HIP-112748, HIP-112051, and HIP-114155 are marked with the first not-saturated star ($V=7.3$) and the dimmest detected star ($V=9.3$). The arrow indicates the directions of the apparent diurnal motion of the stars. The length is the difference in position between this frame and the next one
, which was taken after one minute.
}
\end{figure}

It is worth illustrating the results for a typical night, in our case, Night 23 in Tab.~\ref{fig:night:alt:az:scatter}. 
The observation starts at 2022-01-23T17:01:05 UTC and ends at 2022-01-24T05:54:58 UTC with 774 minutes of observation.
In total, 774 stacked frames have been collected and divided into 15 observing blocks. 
Of them, 624 frames have been accepted. The others were removed since they belonged to too short observing blocks, 
or they were of too bad quality. Frames could be of bad quality due to too short programmed $\texp$,
thermal instabilities, 
image contamination by city lights, 
the presence of clouds, or the lack of a good
GPS signal at the epoch of acquisition.

The accepted frames were processed with \AstrometryNet\ to extract the astrometric information. 
Fig.~\ref{fig:night:frame:id} shows an example of a processed frame. In the figure, circles mark detected stars
in the Tycho 2 catalog. Stars visible to the naked eye are saturated; they are:
HIP-112748 a V=3.6 mag star,
HIP-112051 a V=4.8 mag star,
HIP-114155 a V=4.9 mag star. 
The brightest not saturated star is HIP-112548.0 with V=7.3, while the dimmest star observed with an acceptable S/N is a V=9.3 star with no HIP number.
Stars with V>9.3 could be confused with noise fluctuations.
114 objects are detected with a sufficient S/N for a total $\texptotal \approx 0.1\sec$.

\begin{figure}
\centering
\includegraphics[width=\textwidth]{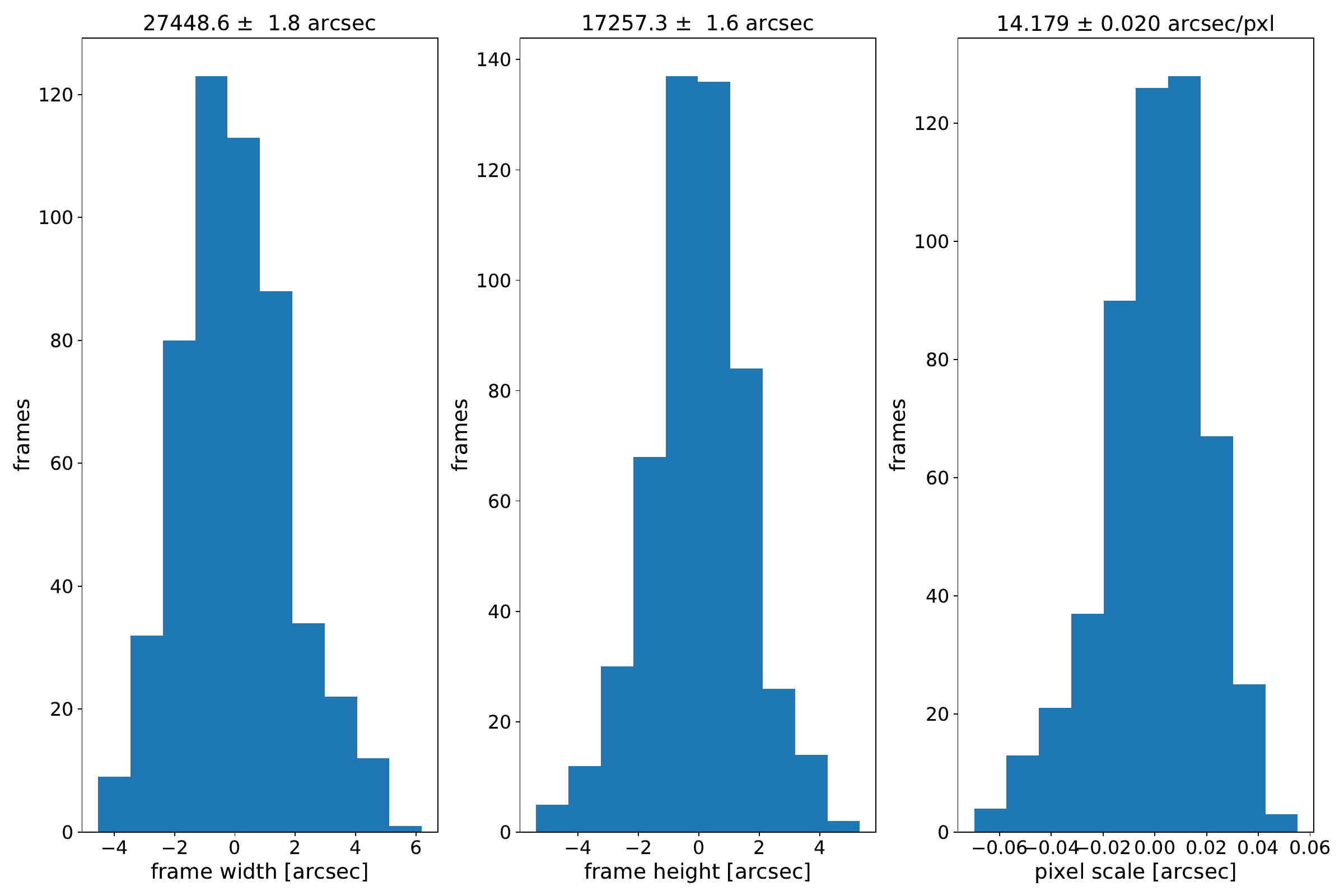}
\caption{\label{fig:night:scales:hist}
Histogram of frame scales for Night 2022 Jan 23rd.
On the abscissa, the mean values of each scale are subtracted. 
}
\end{figure}

Since \AstrometryNet\ performs a blind astrometric reconstruction, for each frame 
an estimate of the pixel scale, width, and height of the frame projected in the sky was provided. 
In some cases, we see significant differences in these parameters with respect to the average. When this happened, the frames were discarded using sigma clipping. 
In this way, 90 more frames were discarded, and the remaining 514 frames were kept for the subsequent data analysis.
Fig.~\ref{fig:night:scales:hist}
presents the histograms of the frame width, frame height, and pixel scale for the 514 frames.
The estimated frame width, height, and pixel scale are respectively $27558.6\pm1.8\;\arcsec$,
$17257.3\pm1.6\;\arcsec$ and $14.178\pm0.020\;\arcsec/\pxl$ which are in line with the design expectations.

\begin{figure}
\centering
\includegraphics[width=\textwidth]{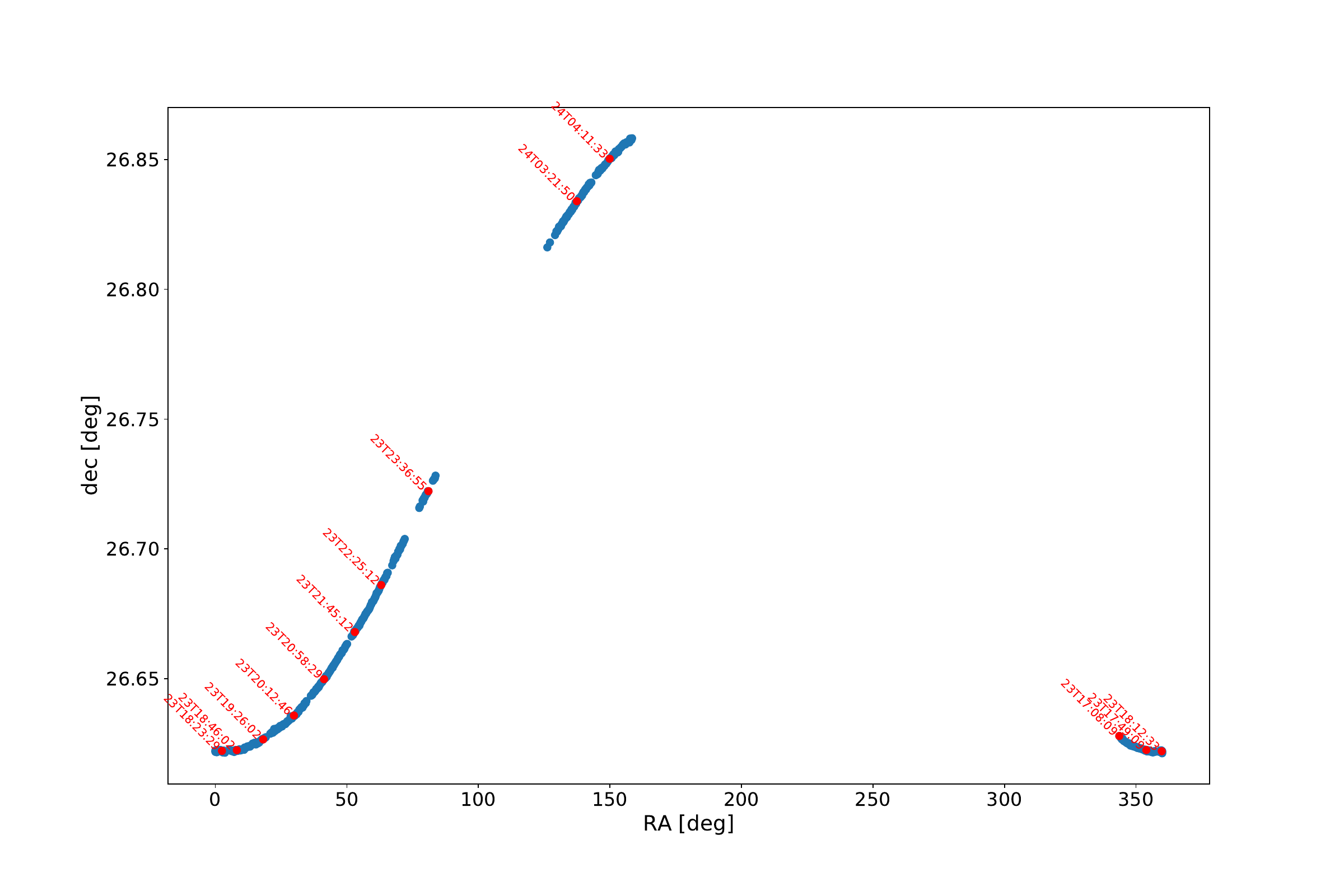}
\caption{\label{fig:night:ra:dec}
J2000 equatorial positions observed on the night of 2022 Jan 23rd.
The UTC day and hour of observations are tagged near the curve.
}
\end{figure}

Fig.~\ref{fig:night:ra:dec}
shows the positions of the constant alt-az pointing directions in the J2000 reference frame as measured from 
\AstrometryNet. 
 The night begins pointing toward $\RA\approx350\deg$, and then the diurnal motion moves to lower values of $\RA$. 
 The interruption between $\RA=100\deg$ and $\RA=150\deg$ is due to a test with frames with too short exposures to detect any star.  
 Declination would have to be constant if expressed in the equinox at the epoch of observation. Still, astronomical and diurnal aberration of light, coupled with the fact that the actual pole of the sky at the epoch of observation is not the J2000 celestial pole, make $\DEC$ a function of time.

We used \astropy{} to convert J2000 equatorial coordinates to apparent alt-az coordinates. They are presented in 
Fig.~\ref{fig:night:alt:az:scatter}, where numbers from 1 to 12 refer to the analyzed observing block.
Ideally, the reconstructed alt-az pointing directions would have to be collapsed onto the same $\Alt$ and $\Az$.
The reconstructed pointing direction for day 23 is $\Alt_{23}=46.38628\deg\pm1.97\arcsec$ and $\Az_{23}=261.95138\deg\pm3.83\arcsec$. 
Not surprisingly the $\Az$ uncertainty is larger than the $\Alt$ uncertainty, however when scaled by $\cos \Alt$ 
the corresponding pointing uncertainty is equivalent to $2.64\;\arcsec$.
Compared to the pixel size $\deltaAlt \approx 0.13\;\pxl$ while $\deltaAz \approx 0.27\;\pxl$. 

Some structure is present in the scatter plot, as it is evident looking at how the sequential numbers are distributed in Fig.~\ref{fig:night:alt:az:scatter}.
A peak to peak variation of $10\;\arcsec$ for $\Az$ and $4\;\arcsec$ for $\Alt$  centroids is seen,
while looking at intra OBs scattering $\deltaAlt < 2.2\;\arcsec$ and $\deltaAz < 1.7\;\arcsec$. 

\begin{table}
\centering
\begin{tabular}{ccccccccccc}
OB & GMT start & $\Delta t$ & \#Fr  & $\texp$ & $\Tair$  &  $\pair$ & $\Alt$  & $\deltaAlt$   & $\Az$    & $\deltaAz$ \\
     &           & [$\sec$] &  & [$\sec$] & $[\deg C]$ & [hpa]             & [deg]& [$\arcsec$]& [deg] & [$\arcsec$] \\
\hline
1 & 01-23T17:08:09 & 3300 & 55 & 0.140 & 6.6 & 1019.3 & 46.3860 & 1.05 & 261.9525 & 1.00\\
2 & 01-23T18:08:09 &  456 & 58 & 0.105 & 6.5 & 1019.5 & 46.3865 & 0.84 & 261.9528 & 1.06\\
3 & 01-23T18:20:49 &  450 & 37 & 0.105 & 6.3 & 1019.5 & 46.3865 & 1.01 & 261.9527 & 1.01\\
4 & 01-23T18:35:02 & 3300 & 55 & 0.105 & 6.1 & 1019.6 & 46.3863 & 1.16 & 261.9519 & 1.73\\
5 & 01-23T19:36:46 & 3300 & 55 & 0.070 & 5.4 & 1019.5 & 46.3858 & 2.16 & 261.9515 & 1.43\\
6 & 01-23T20:38:29 & 3300 & 56 & 0.105 & 5.2 & 1019.5 & 46.3858 & 1.30 & 261.9511 & 0.96\\
7 & 01-23T21:40:12 & 3300 & 55 & 0.140 & 4.9 & 1019.7 & 46.3860 & 1.28 & 261.9504 & 1.13\\
8 & 01-23T22:41:55 & 3300 & 30 & 0.105 & 5.0 & 1020.0 & 46.3863 & 1.80 & 261.9506 & 1.55\\
9 & 01-23T23:43:39 &  240 & 5 & 0.070 & 4.9 & 1019.9 & 46.3860 & 1.00 & 261.9504 & 1.43\\
12 & 01-24T02:48:50 & 3300 & 55 & 0.105 & 5.6 & 1020.0 & 46.3869 & 1.27 & 261.9502 & 1.32\\
13 & 01-24T03:50:33 & 3300 & 51 & 0.070 & 5.3 & 1020.0 & 46.3868 & 1.39 & 261.9501 & 1.62\\
\end{tabular}
\caption{\label{tab:night:observing:blocks}
Table of observing blocks (OB) in Fig.~\ref{fig:night:alt:az:scatter}. 
Different experiments correspond to different $\texp$.
The GMT start refers to the first valid frame, the same for the time interval spanned by each run $\Delta t$ and the number of frames collected.
For all the observing blocks, all the stacked frames have acquired with a 60\,s cadence, apart from blocks 2 and 3, for which a 10~$\sec$ cadence has been chosen.
}
\end{table}

\begin{figure}
\centering
\includegraphics[width=\textwidth]{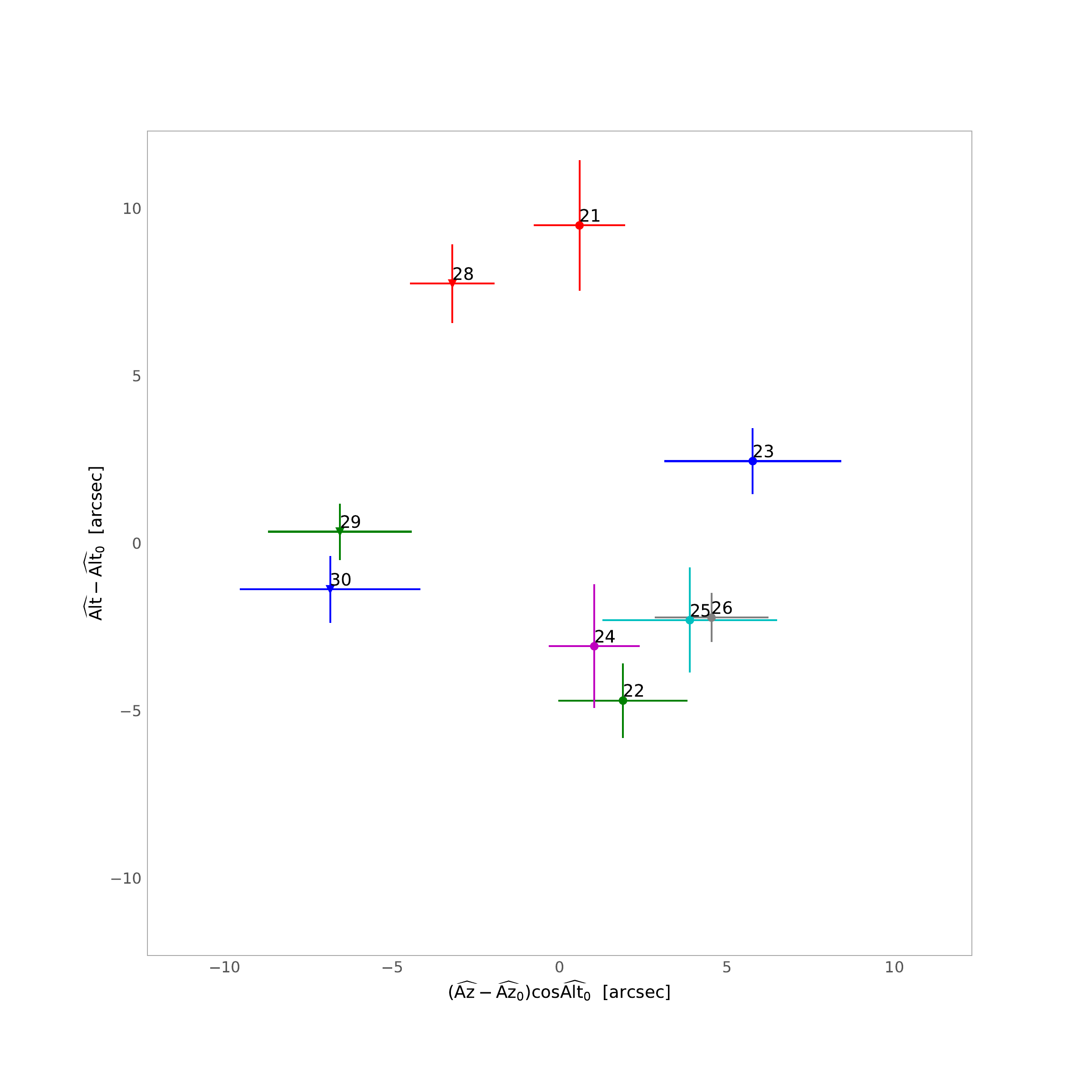}
\caption{\label{fig:west:night:alt:az}
Reconstructed alt-az centroids for nights between Jan 21st and Jan 30th, 2022.
Numbers refer to the different nights in Tab.~\ref{tab:nights:observations}. 
The plot is centered on the weighted average of the nine nights 
$\AltZ= 46.38559 \deg$ 
and 
$\AzZ= 261.94906 \deg$. 
 Azimuth deviations and error bars are scaled by $\cos \AltZ$.
}
\end{figure}

Tab.~\ref{tab:nights:observations} gives the night-by-night result for the whole set of observations in the test; the reconstructed 
alt-az coordinates are shown in Fig.~\ref{fig:west:night:alt:az}.
The plot is centered on the weighted average of the nine nights: $\AltZ= 46.38559 \deg$ and $\AzZ= 261.94906\deg$. 
Error bars in the figure for $\Az$ and the deviation from the night mean $(\Alt - \AltZ)$ are scaled by  $\cos \AltZ$.
It is evident how night by night, there is some shift of the centroids larger than the night RMS,
but a clear pattern does not emerge.
The centroids cluster into four groups: nights 21 and 28, nights 23, nights 25 and 30, and nights 22, 24, 25, and 26.
The centre-to-centre distance between the clusters in $\Az$ is $\approx 18\;\arcsec$ equivalent to a pointing difference of $\approx 12\;\arcsec$. 
The centre-to-centre distance between the clusters in $\Alt$ is $\approx 12\;\arcsec$.

\begin{figure}
\centering
\includegraphics[width=\textwidth]{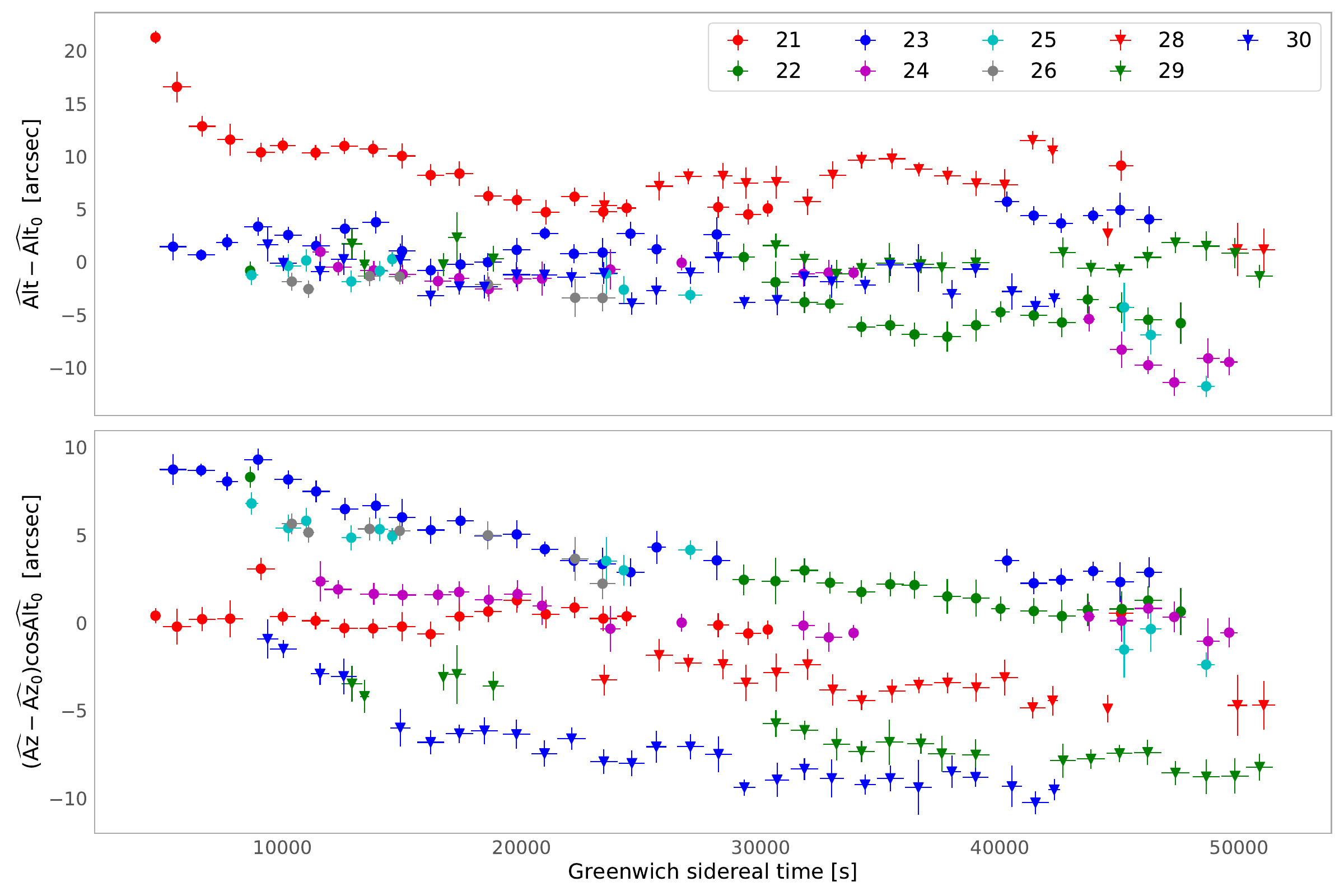}
\caption{\label{fig:west:centroided:alt:az:vz:st}
Deviations from $\AltZ$ and $\AzZ$ for nights between Jan 21st and Jan 30th, 2022, as a function of sidereal time. 
Data from each night are binned in sidereal time intervals of 1200$\sec$.
}
\end{figure}

A residual drift during the night can be seen in Fig.~\ref{fig:west:centroided:alt:az:vz:st} showing the $\Alt$ and $\Az$ as a function of 
sidereal time. Data belonging to the same night are binned in sidereal time intervals of 1200$\sec$\footnote{Tests with other time intervals give similar results.}, 
so that the circles are the averaged values in each bin while the error bars are the standard
deviations of each bin.
Looking at $\Alt$ v.z. sidereal time, nights 21 and 28 are separated from the other nights.
A drift of at most 10 arcsec across the night is apparent. The $\Az$ v.z. sidereal time has a larger dispersion, but some drift can also be seen more or less with the same amplitude as in $\Alt$. However, looking at the individual nights, it is difficult to derive a coherent behavior.

More interesting is the small dispersion within each bin, which may represent an estimate of the random uncertainty in a single measure. 
The distribution in time of the standard deviation for binned $\Alt$ and $\Az$ are not correlated with time or any other relevant parameter.
In 95\% of the cases $\Alt$ bins have a standard deviation of less than $1.9\,\arcsec$ while the corresponding percentile in $\Az$ bins (scaled by $\cos \AltZ$) is $\approx1.5\;\arcsec$,
so in most of the cases, the random pointing error has to be less than $3\,\arcsec$

In conclusion, from those tests, the prototype of the STR has a random accuracy of $3\;\arcsec$ and an upper limit instability of approx $10\;\arcsec$.
It has to be noted that given the physical size of the device, a pointing shift of $10\,\arcsec$ is equivalent to a relative shift of the various parts of the prototype of 
at most $14\,\micron$. Given the conditions of the test, it is not possible to exclude that in ten days, 
the various parts of the STR or the mount itself
did not move of a similar amount due either to the wind pressure, the application of the cover protecting the instrument during the day, and so on. 
The better mechanical setup planned for the final version of the STR is likely to reduce the drift significantly.

It is interesting to compare this result with a test conducted in June 2021 while debugging the STR onboard software.
In this test, we observed at night the lights of a group of large antennas
located near the village of {\em Banne}, on the top of an hill, about $3.8\;\Km$ north-east of the OATs position.
We observed for six nights in 2021: June 22, 23, 26, 27, and 28. We fixed the STR during the test.
We used the average of the positions of the lights in each frame to determine the position
of the frame center with respect to the trelly; Its
RMS represents a measure of the accuracy of geometric pointing.
Assuming the averaged pixel scale  $14.2\;\arcsec/\pxl$ the result is an RMS between 
$2.2\;\arcsec$ and $4\;\arcsec$ compatible with the results of the sky observations.
Depending on the night, a residual time drift at $5\;\arcsec$ was seen in the data, not too far from what was observed in the sky.
Note that buildings did not impede the line of sight from our station to the antennas, but it was a few degrees above the horizon and
crossed a large part of the air column above the city, resulting in significant image turbulence. Also, a $5\;\arcsec$ drift
is equivalent to a $9\;\cm$ relative change in the position of the antenna, which is compatible with the effect of Earth tides
or flexures in the trelly due to wind pressure or changes in the air temperature.


\subsection{Building a pointing model from Star Tracker observations}\label{sec:str:prm}

We consider here the accuracy attainable for determining a pointing model using the STR as a case study. 
For a given list of $\nsimp$ combinations of control angles $(\EncoderAlt,\EncoderAz)_i,$ ($i=0,1, \dots \nsimp-1)$ we simulated the corresponding pointings measurements $\PointingSi$ obtained by the STR in the Geo reference frame. 
Here we consider a grid of 40 observations, equivalent to 5 nights of operations,  distributed over $\ntheta=5$ circles of constant $\EncoderAlt$ in the evenly sampled interval $[65\deg,75\deg]$ ($[25\deg,15\deg]$ of zenith distance),
and $\nphi=8$ evenly sampled $\EncoderAz$ values covering the range $[0\deg,360\deg)$.
A $\PointingSi$ is the sum of the geometrical pointing provided by the PRM and an STR random pointing error.
The geometrical pointing is generated for a given set of configuration angles
$\mathbf{\Theta}_{\mathrm{fit}}=(\omegaVAX,\zVAX,\varphiRef,\tiltFork,\varthetaRef)$ which parameterized the simulation, to simplify the test, we neglect the camera configuration angles.
We model the STR random pointing error on the results of the STR prototype, so the error is the sum of a bidimensional Gaussian noise with $\sigma = 3\;\arcsec$ and a uniform bidimensional circular top hat distribution with radius $\strdrift=10\;\arcsec$. The circular top hat is the dominant source of error, and it represents the worst case for the drift seen in the pointing reconstructed from the STR during the testing campaign.

The pointing model to be obtained from the data has to provide the corresponding list of $\PointingMi$.
A possible approach to determine such model is to use $\PointingSi$ to recover   $(\omegaVAX,\zVAX,\varphiRef,\tiltFork,\varthetaRef)$ by minimizing

\begin{equation}\label{eq:str:prm:chisq}
    \chi^2(\mathbf{\Theta}_{\mathrm{fit}}) = \sum_{i=0}^{\nsimp-1} \left| \PointingMi(\mathbf{\Theta}_{\mathrm{fit}}) -\PointingSi \right|^2.
\end{equation}

\noindent
However, this approach has some drawbacks, particularly for small values of the configuration angles, where some level of degeneracy occurs.

A blind method could be implemented by noting from Eq.~(\ref{eq:AttitudeMatrix}) that for evenly sampled $\EncoderAz$ and constant $\EncoderAlt$ the pointing direction averaged over a circle
$\PointCaver=\sum_i \PointingSi/\nphi$ 
is parallel to the V-AXIS\footnote{To demonstrate it replace $\rotZ(\EncoderAz)$ in Eq.~(\ref{eq:AttitudeMatrix}) with its average.
}. 
It is also possible to define the cosine transform for the circle $\PointCcos = \sum_i \PointingSi \cos \EncoderAzi/\sum_i\cos^2 \EncoderAzi$ and similarly, the sine transform $\PointCsin$, in this way the pointing model for a given circle is 

\begin{equation}\label{eq:pmodel:fft}
    \Pointing_c(\EncoderAz) = \PointCaver + \PointCcos \cos \EncoderAz +  \PointCsin \sin \EncoderAz. 
\end{equation}

\noindent
For other values of $\EncoderAlt$, the corresponding vectors can be derived by quadratic interpolation. 
Using the pointing error defined in Eq.~(\ref{eq:deflP}), we obtain an average pointing accuracy of $4\;\arcsec \pm 2\;\arcsec$.

As noted before, the $\PointCaver$ are parallel to V-AXIS so that $\zVAX$ and $\omegaVAX$ can be derived directly from those vectors.
The other configuration angles $(\varphiRef,\tiltFork,\varthetaRef)$ can be derived from $\chi^2$ minimization with less numerical instability.
We tested the method for different combinations of PRM parameters. 
As expected, $\zVAX = 0\deg$ for all cases, and the fitted value of $\omegaVAX$ spans its whole range.
The other angles are recovered with an accuracy of
$1\;\arcsec$ for $\zVAX$, 
$11\;\arcsec$ for $\varphiRef$,
$4\;\arcsec$ for $\tiltFork$,
$1\;\arcsec$ for $\varthetaRef$.
When $\zVAX \ne 0$ the typical accuracy for the determination of $\omegaVAX$ is $\approx12\;\arcmin$.
The estimators for all of the angles are mutually uncorrelated apart from the case of 
$\varphiRef$ and $\tiltFork$, which has a correlation coefficient of $-0.98$. 
Pointing errors can be decomposed into a co-scan and a cross-scan component. The first is the pointing error along the scan direction, and the other is the pointing error that is normal to it. 
The mean of co-scan and cross-scan errors is zero with RMS $1\;\arcsec$ and $1.1\;\arcsec$, respectively. Adding in squares the two errors is equivalent to computing the deflection of Eq.~(\ref{eq:deflP}), resulting in a mean deflection of $1.3\;\arcsec$. For 95\% of the cases, the error is below $2.6\;\arcsec$. 
By propagating those uncertainties with Eq.~(\ref{eq:distribution:scaling}) and using the sum in squares rule, we predict $\gamma_P \lesssim 3 \times 10^{-3} \; \microK$ which represents an upper limit for the effect of the pointing error given we assumed the uncertainties on the control angles to be uncorrelated. 

As a last remark, in this study, the PRM method outperformed the blind method, given its ability to combine more observations with fewer free parameters.
However, before concluding that the pointing reconstruction based on the PRM is better than the blind method, it has to be taken into account that the PRM presented in Sect.~\ref{sec:PRM} assumes a perfectly stiff telescope which is not the case for \STRIPLSPE. In particular, when looking at different altitudes, changes in flexures could result in pointing errors not described by the current version of the PRM but that can be mapped by the blind method. 


\subsection{Intercalibration}\label{sec:prm:intercalibration}

The accuracy of the pointing quoted above is relative to the use of the PRM to point the STR.
Of course, even in the ideal conditions of no flexures, there will be some angular offset between the STR l.o.s. and the telescope l.o.s. $\strTelOffset$. 
An intercalibration method to measure $\strTelOffset$ should be provided to complete the calibration of the PRM.
Two possible intercalibration strategies are available. In the first, $\strTelOffset$ is measured by using a drone (UAV) equipped with a radio beacon and a bright LED; in the second, planets or other very bright radio sources will be used. 

We begin our discussion by focusing on UAV-based calibration. In recent years, UAV-based measurement campaigns have gained significant attention for the testing and calibration of modern radio telescopes \cite{drone:paonessa:etal:ska,drone:ciorba:etal,drone:virone:etal:lofarA,drone:dininni:etal:lofarB}. Continuous wave (CW) transmitters have been utilized across a wide frequency range, from the VHF band to the Q-band \cite{drone:paonessa:etal:2020}. In our case, a UAV equipped with a narrowband radio transmitter delivers sufficient power to achieve detection by \STRIPLSPE\ with a remarkably high signal-to-noise ratio (S/N), thereby enabling the calibration of the instrument’s beam pattern.
 %
 %
In the baseline mode of operations, the UAV will fly at about $120\;\meter$ from the telescope, rastering an area of sky in front of it. 
The relative position of the UAV with respect to the telescope will be determined within $2\;\cm$~rms by using a differential GPS (dGPS).
To allow the STR to properly identify the drone, the UAV will be equipped with a very bright LED located in a known position with respect to the dGPS antenna. 
The angular offset between the STR and the central beam of \STRIP\ can be determined by comparing the angular position of the centroid of the beam with the angular position of the center of the scanning area, as seen from the STR.
In a preliminary test performed in July 2018 with a previous STR prototype, we were able to measure the position of the center with $6\,\arcsec$ RMS
\cite{report:drone:test:2018}.
However given the UAV is flying at a finite distance from \STRIP\ we need to account for parallaxes. This means including corrections for the relative position of the LED and the radio beacon with respect to the dGPS antenna onboard the UAV, the relative position of the STR with respect to the optical axis of the telescope, the relative position of the telescope with respect to the origin of the dGPS reference frame. 
The RMS positioning accuracy of the UAV provided by the dGPS is equivalent to $34.4\;\arcsec$; during the test, we have been able to acquire up to $3\times10^3$ positions with the STR, so the averaged angular positional uncertainty will be $\approx1\;\arcsec$.
By construction, the relative position of the LED and the radio beacon with respect to the dGPS antenna is $0.5\;\cm$ equivalent to an uncertainty of $\approx9\;\arcsec$ at $120\;\meter$ of distance. The STR is located at $\approx1\;\meter$ from the optical axis of the telescope, and the accuracy by which its position can be determined is $\approx1\div2\;\cm$ equivalent to $17\div34\;\arcsec$ of systematic offset. 
The error of positioning of the origin of the UAV dGPS with respect to the optical axis and the STR largely depends on how the dGPS system is implemented, and it varies between $0.1 \div 2 \; \cm$. 
Therefore, the uncertainty in the parallactic correction varies from $0.25\;\arcmin$ to $1\;\arcmin$, which represents the main limitation of this method.
In principle, improving this number by flying the UAV at a larger distance would be possible. However, this must consider aerial regulations imposing a maximum height of $120\;\meter$ for the UAV.
A second possibility would be to repeat the raster scan at different distances to measure the variation of the parallactic effect. 
Implementing such a multi-distance scan is under evaluation and will be the subject of future work.

The two main limitations in observing a planet are the epoch of observability and the SN ratio. First, a planet like Jupiter will be observable about twice a year. Estimating the S/N ratio is more complicated as planets must be observed in Total Power rather than Polarization.  
For the Q-band expected NET from \cite{2021JCAP...08..008A}, the peak S/N ratio, which can be expected for a planet like Jupiter, is $\approx5\times10^2$. 
However, reliable laboratory tests for the radiometer performances are available only for polarization. No clear indications for the $1/f$ noise knee frequency are available for observations in total power, so it is expected that due to the $1/f$ noise, the final noise in Total Power will be much higher than in Polarization. In addition, complications such as the atmosphere, beam smearing, and so on will further increase the confusion noise.
The high brightness of a planet like Jupiter allows us to parametrize a detection model using a peak S/N ratio. Here, we consider two optimistic cases: $S/N=10$ and $S/N=50$ for Jupiter observed within the beam of the central horn. 
Given the need of the STR to have stable images, not blurred by the telescope motion, we assumed to raster the planet in a rectangular grid of $5\times5$ fixed positions centered around the planet with $40\;\arcmin$ side. Assuming to take $1~\min$ of observation and $1~\min$ for repointing, we expect to complete a raster in one hour. A short time allows the raster to be performed near the upper culmination and to repeat it at different epochs to reduce the impact of atmospheric instability. With those parameters, the accuracy by which Jupiter can be positioned within the STR is $\approx 10\;\arcsec$. The accuracy by which the beam center can be determined by the radio observations of Jupiter depends on the $S/N$ ratio. So for the $S/N=50$ case we estimate an accuracy in determining $\strTelOffset$ of $\approx 1/3\;\arcmin$, for the $S/N=10$ case the accuracy is $\approx 1\;\arcmin$ 
with the accuracy which scales approximately as $(S/N)^{3/4}$.

In short, the final pointing accuracy will be dominated by the accuracy by which $\strTelOffset$ will be determined, but we expect to be able to keep this error within $\approx1\;\arcmin$.


\section{Conclusions and future work}\label{sec:conclusions}

This paper presents the first Pointing Reconstruction Model (PRM) for the \STRIPLSPE\ telescope. The model is at the root of the pointing reconstruction pipeline of the instrument and it is also part of the simulation pipeline. 
The PRM was used to test the effect of uncertainties in the telescope configuration, represented by a set of configuration angles, leading to uncertainties in the pointing reconstruction. 

The \STRIPLSPE\ baseline fixes a pointing accuracy of $30\,\arcsec$, a demanding requirement for telescope mechanics. 
The baseline was defined from simple assumptions about the effect of pointing error, and no detailed study was provided. 
Here, we used the PRM to simulate the error in the polarization maps produced by \STRIPLSPE\ in different scenarios of telescope pointing accuracy, including the baseline. We show a linear increase in the rms of the noise of the polarization map due to the pointing error, which remains one or two orders of magnitude below the effect of the instrumental noise. 
This allows us to propose a more feasible baseline for the \STRIPLSPE\ pointing accuracy of $1\;\arcmin$. 

\STRIPLSPE\ will need a Star Tracker (STR) to calibrate the PRM against optical astronomical observations. In this work, we presented the design of the prototype  STR for \STRIPLSPE\ and the results of a testing campaign aimed at characterizing its accuracy. The prototype shows a random pointing error better than $3\;\arcsec$ and a long-term variability of $\approx10\;\arcsec$. 
It is likely that the long-term variability is mainly caused by a lack of stiffness in the mechanical support hosting the prototype; if so, the final version of the STR will be more stable.
We used the results of the testing campaign to simulate the calibration of a pointing model. Our analysis shows that with the current configuration, the average expected pointing error would be better than $4\;\arcsec$.  

The next step is to implement a more detailed PRM that includes encoder errors, flexures, thermal effects on the telescope mechanics, and other pointing perturbations.
In addition, we will analyze the geometrical intercalibration of the STR and the radio telescope. We plan to use a drone carrying a radio and a light source, and we foresee the observation of bright radio sources, such as Jupiter or Venus.

\acknowledgments
The LSPE/Strip Project is carried out thanks to the support of ASI (contract I/022/11/1 and Agreement 2018-21-HH.0) and INFN. The Strip instrument will be installed at the Observatorio del Teide, Tenerife, of the Instituto de Astrofísica de Canarias.



This work has been partially funded by project AYA2017-84185-P of the Spanish Ministry of Science and Innovation (MICINN).


The Astronomical Observatory of the Autonomous Region of the Aosta Valley (OAVdA) is managed by the Fondazione Clément Fillietroz-ONLUS, which is supported by the Regional Government of the Aosta Valley, the Town Municipality of Nus and the "Unité des Communes valdôtaines Mont-Émilius". Stefano Sartor acknowledges funds from a 2020 ‘Research and Education’ grant from Fondazione CRT-Cassa di Risparmio di Torino.

In response to the COVID-19 pandemic, which severely limited the activities of the INAF/OATs laboratories, other organizations helped us set up the Star Tracker prototype. We particularly acknowledge the Mechanical Workshop of Milano University's Physics department, the Trieste ICTP/Scientific FabLab and the Trieste MG7Maker-Lab.

In addition, MM acknowledges the technical support provided by Silvio Burolo, Federico Gasparo, Bertocco Sara, Igor Coretti, and Fabio Stocco of the IT division of Trieste Astronomical Observatory in setting up the testing facility.

The preliminary analysis for this project was part of a one-month internship for undergraduate students at OATs in 2018. MM wishes to acknowledge the work of Matteo Porru and Yousef El Sharkawy.

Also, Matteo Di~Mario did some preliminary analysis of the role of flexures in pointing errors in his Master's Thesis. As the subject of flexures is not within the scope of this paper, those results will be the subject of dedicated work in the future.

This research made use of
{\tt Arduino cli}\footnote{\url{https://www.arduino.cc/pro/cli}};
{\tt Arduino IDE}\footnote{\url{https://github.com/arduino/Arduino}};
{\tt ASI Camera Boost} a free library from Pawel Soya\footnote{\url{https://github.com/pawel-soja/AsiCamera}};
{\tt Astropy}\footnote{\url{http://www.astropy.org}} a community-developed core \Python\ package for astronomy \citep{astropy:2013, astropy:2018};
{\tt Asymptote}\footnote{\url{https://asymptote.sourceforge.io/}} \citep{asymptote:2010};
{\tt EMCEE}\footnote{\url{https://github.com/dfm/emcee}} \citep{emcee};
{\tt HEALPix}\footnote{\url{https://healpix.jpl.nasa.gov}} \citep{Healpix};
{\tt HEALPy}\footnote{\url{https://github.com/healpy/healpy}} \citep{Healpy};
{\tt IPython}\footnote{\url{https://ipython.org}} \citep{PER-GRA:2007};
{\tt MatPlotLib}\footnote{\url{https://matplotlib.org}} \citep{matplotlib};
{\tt NumPy}\footnote{\url{https://numpy.org}} \citep{2020SciPy-NMeth};
{\tt Pandas}\footnote{\url{https://pandas.pydata.org}} \citep{mckinney-proc-scipy-2010}.
{\tt PyQuarantine}\footnote{\url{https://www.ict.inaf.it/gitlab/michele.maris/pyquarantine.git}};
%
{\tt SciPy}\footnote{\url{https://www.scipy.org}} \citep{2020SciPy-NMeth};
{\tt YAPSUT}\footnote{\url{https://www.ict.inaf.it/gitlab/michele.maris/yapsut.git}}. 
%

\bibliographystyle{JHEP}
\bibliography{H10_PR_STR}

\newcommand{\noop}[1]{}

\providecommand{\href}[2]{#2}\begingroup\raggedright\begin{thebibliography}{10}

\bibitem{2016ASSL..423.....M}
\emph{{Understanding the Epoch of Cosmic Reionization}}, vol.~423 of
  \emph{Astrophysics and Space Science Library}, Jan., 2016.
\newblock 10.1007/978-3-319-21957-8.

\bibitem{2013ApJS..208...19H}
G.~{Hinshaw}, D.~{Larson}, E.~{Komatsu}, D.N.~{Spergel}, C.L.~{Bennett},
  J.~{Dunkley} et~al., \emph{{Nine-year Wilkinson Microwave Anisotropy Probe
  (WMAP) Observations: Cosmological Parameter Results}},
  \href{https://doi.org/10.1088/0067-0049/208/2/19}{\emph{\apjs} {\bfseries
  208} (2013) 19} [\href{https://arxiv.org/abs/1212.5226}{{\ttfamily
  1212.5226}}].

\bibitem{2016A&A...596A.108P}
{Planck Collaboration}, R.~{Adam}, N.~{Aghanim}, M.~{Ashdown}, J.~{Aumont},
  C.~{Baccigalupi} et~al., \emph{{Planck intermediate results. XLVII. Planck
  constraints on reionization history}},
  \href{https://doi.org/10.1051/0004-6361/201628897}{\emph{\aap} {\bfseries
  596} (2016) A108} [\href{https://arxiv.org/abs/1605.03507}{{\ttfamily
  1605.03507}}].

\bibitem{1997PhRvL..78.2054}
U.~{Seljak} and M.~{Zaldarriaga}, \emph{{Signature of Gravity Waves in the
  Polarization of the Microwave Background}},
  \href{https://doi.org/10.1103/PhysRevLett.78.2054}{\emph{\prl} {\bfseries 78}
  (1997) 2054} [\href{https://arxiv.org/abs/astro-ph/9609169}{{\ttfamily
  astro-ph/9609169}}].

\bibitem{1997PhRvL..78.2058K}
M.~{Kamionkowski}, A.~{Kosowsky} and A.~{Stebbins}, \emph{{A Probe of
  Primordial Gravity Waves and Vorticity}},
  \href{https://doi.org/10.1103/PhysRevLett.78.2058}{\emph{\prl} {\bfseries 78}
  (1997) 2058} [\href{https://arxiv.org/abs/astro-ph/9609132}{{\ttfamily
  astro-ph/9609132}}].

\bibitem{1997PhRvD..56..596H}
W.~{Hu} and M.~{White}, \emph{{CMB anisotropies: Total angular momentum
  method}}, \href{https://doi.org/10.1103/PhysRevD.56.596}{\emph{\prd}
  {\bfseries 56} (1997) 596}
  [\href{https://arxiv.org/abs/astro-ph/9702170}{{\ttfamily
  astro-ph/9702170}}].

\bibitem{2019JCAP...02..056A}
P.~{Ade}, J.~{Aguirre}, Z.~{Ahmed}, S.~{Aiola}, A.~{Ali}, D.~{Alonso} et~al.,
  \emph{{The Simons Observatory: science goals and forecasts}},
  \href{https://doi.org/10.1088/1475-7516/2019/02/056}{\emph{\jcap} {\bfseries
  2019} (2019) 056} [\href{https://arxiv.org/abs/1808.07445}{{\ttfamily
  1808.07445}}].

\bibitem{2022JCAP...04..034H}
J.C.~{Hamilton}, L.~{Mousset}, E.S.~{Battistelli}, P.~{de Bernardis},
  M.A.~{Bigot-Sazy}, P.~{Chanial} et~al., \emph{{QUBIC I: Overview and science
  program}}, \href{https://doi.org/10.1088/1475-7516/2022/04/034}{\emph{\jcap}
  {\bfseries 2022} (2022) 034}
  [\href{https://arxiv.org/abs/2011.02213}{{\ttfamily 2011.02213}}].

\bibitem{2022ApJ...926...54A}
K.~{Abazajian}, G.E.~{Addison}, P.~{Adshead}, Z.~{Ahmed}, D.~{Akerib}, A.~{Ali}
  et~al., \emph{{CMB-S4: Forecasting Constraints on Primordial Gravitational
  Waves}}, \href{https://doi.org/10.3847/1538-4357/ac1596}{\emph{\apj}
  {\bfseries 926} (2022) 54}
  [\href{https://arxiv.org/abs/2008.12619}{{\ttfamily 2008.12619}}].

\bibitem{2023PTEP.2023d2F01L}
{LiteBIRD Collaboration}, E.~{Allys}, K.~{Arnold}, J.~{Aumont}, R.~{Aurlien},
  S.~{Azzoni} et~al., \emph{{Probing cosmic inflation with the LiteBIRD cosmic
  microwave background polarization survey}},
  \href{https://doi.org/10.1093/ptep/ptac150}{\emph{Progress of Theoretical and
  Experimental Physics} {\bfseries 2023} (2023) 042F01}
  [\href{https://arxiv.org/abs/2202.02773}{{\ttfamily 2202.02773}}].

\bibitem{2021PhRvL.127o1301A}
P.A.R.~{Ade}, Z.~{Ahmed}, M.~{Amiri}, D.~{Barkats}, R.B.~{Thakur},
  C.A.~{Bischoff} et~al., \emph{{Improved Constraints on Primordial
  Gravitational Waves using Planck, WMAP, and BICEP/Keck Observations through
  the 2018 Observing Season}},
  \href{https://doi.org/10.1103/PhysRevLett.127.151301}{\emph{\prl} {\bfseries
  127} (2021) 151301} [\href{https://arxiv.org/abs/2110.00483}{{\ttfamily
  2110.00483}}].

\bibitem{2020A&A...643A..42P}
{Planck Collaboration}, Y.~{Akrami}, K.J.~{Andersen}, M.~{Ashdown},
  C.~{Baccigalupi}, M.~{Ballardini} et~al., \emph{{Planck intermediate results.
  LVII. Joint Planck LFI and HFI data processing}},
  \href{https://doi.org/10.1051/0004-6361/202038073}{\emph{\aap} {\bfseries
  643} (2020) A42} [\href{https://arxiv.org/abs/2007.04997}{{\ttfamily
  2007.04997}}].

\bibitem{2022PhRvD.105h3524T}
M.~{Tristram}, A.J.~{Banday}, K.M.~{G{\'o}rski}, R.~{Keskitalo},
  C.R.~{Lawrence}, K.J.~{Andersen} et~al., \emph{{Improved limits on the
  tensor-to-scalar ratio using BICEP and Planck data}},
  \href{https://doi.org/10.1103/PhysRevD.105.083524}{\emph{\prd} {\bfseries
  105} (2022) 083524} [\href{https://arxiv.org/abs/2112.07961}{{\ttfamily
  2112.07961}}].

\bibitem{2021JCAP...08..008A}
G.~{Addamo}, P.A.R.~{Ade}, C.~{Baccigalupi}, A.M.~{Baldini}, P.M.~{Battaglia},
  E.S.~{Battistelli} et~al., \emph{{The large scale polarization explorer
  (LSPE) for CMB measurements: performance forecast}},
  \href{https://doi.org/10.1088/1475-7516/2021/08/008}{\emph{\jcap} {\bfseries
  2021} (2021) 008} [\href{https://arxiv.org/abs/2008.11049}{{\ttfamily
  2008.11049}}].

\bibitem{2023MNRAS.519.3383R}
J.A.~{Rubi{\~n}o-Mart{\'\i}n}, F.~{Guidi}, R.T.~{G{\'e}nova-Santos},
  S.E.~{Harper}, D.~{Herranz}, R.J.~{Hoyland} et~al., \emph{{QUIJOTE scientific
  results - IV. A northern sky survey in intensity and polarization at 10-20
  GHz with the multifrequency instrument}},
  \href{https://doi.org/10.1093/mnras/stac3439}{\emph{\mnras} {\bfseries 519}
  (2023) 3383} [\href{https://arxiv.org/abs/2301.05113}{{\ttfamily
  2301.05113}}].

\bibitem{lspebeams2022}
S.~{Realini}, C.~{Franceschet}, F.~{Villa}, M.~{Sandri}, G.~{Addamo},
  P.~{Alonso-Arias} et~al., \emph{{The LSPE-Strip beams}},
  \href{https://doi.org/10.1088/1748-0221/17/01/P01028}{\emph{Journal of
  Instrumentation} {\bfseries 17} (2022) P01028}
  [\href{https://arxiv.org/abs/2109.01440}{{\ttfamily 2109.01440}}].

\bibitem{2022JInst..17P1029F}
C.~{Franceschet}, F.~{Del Torto}, F.~{Villa}, S.~{Realini}, R.~{Bongiolatti},
  O.A.~{Peverini} et~al., \emph{{The LSPE-Strip feed horn array}},
  \href{https://doi.org/10.1088/1748-0221/17/01/P01029}{\emph{Journal of
  Instrumentation} {\bfseries 17} (2022) P01029}
  [\href{https://arxiv.org/abs/2107.13775}{{\ttfamily 2107.13775}}].

\bibitem{2022JInst..17P6042P}
O.A.~{Peverini}, M.~{Lumia}, Z.~{Farooqui}, G.~{Addamo}, G.~{Virone},
  F.~{Paonessa} et~al., \emph{{Q-band polarizers for the LSPE-Strip correlation
  radiometric instrument}},
  \href{https://doi.org/10.1088/1748-0221/17/06/P06042}{\emph{Journal of
  Instrumentation} {\bfseries 17} (2022) P06042}.

\bibitem{Genova-Santos:etal:2023}
R.T.~{G{\'e}nova-Santos}, M.~{Bersanelli}, C.~{Franceschet}, M.~{Gervasi},
  C.~{L{\'o}pez-Caraballo}, L.~{Mandelli} et~al., \emph{{LSPE-STRIP on-sky
  calibration strategy using bright celestial sources}},
  \href{https://doi.org/10.48550/arXiv.2401.03802}{\emph{arXiv e-prints} (2024)
  arXiv:2401.03802} [\href{https://arxiv.org/abs/2401.03802}{{\ttfamily
  2401.03802}}].

\bibitem{thelspecollaboration2020large}
G.~{Addamo}, P.A.R.~{Ade}, C.~{Baccigalupi}, A.M.~{Baldini}, P.M.~{Battaglia},
  E.S.~{Battistelli} et~al., \emph{{The large scale polarization explorer
  (LSPE) for CMB measurements: performance forecast}},
  \href{https://doi.org/10.1088/1475-7516/2021/08/008}{\emph{\jcap} {\bfseries
  2021} (2021) 008} [\href{https://arxiv.org/abs/2008.11049}{{\ttfamily
  2008.11049}}].

\bibitem{Zonca2021}
A.~Zonca, B.~Thorne, N.~Krachmalnicoff and J.~Borrill, \emph{The python sky
  model 3 software}, \href{https://doi.org/10.21105/joss.03783}{\emph{Journal
  of Open Source Software} {\bfseries 6} (2021) 3783}.

\bibitem{healpix_jl}
M.~{Tomasi} and Z.~{Li}, \emph{{Healpix.jl: Julia-only port of the HEALPix
  library}},  Sept., 2021.

\bibitem{Healpix}
K.M.~{G{\'o}rski}, E.~{Hivon}, A.J.~{Banday}, B.D.~{Wand elt}, F.K.~{Hansen},
  M.~{Reinecke} et~al., \emph{{HEALPix: A Framework for High-Resolution
  Discretization and Fast Analysis of Data Distributed on the Sphere}},
  \href{https://doi.org/10.1086/427976}{\emph{\apj} {\bfseries 622} (2005) 759}
  [\href{https://arxiv.org/abs/astro-ph/0409513}{{\ttfamily
  astro-ph/0409513}}].

\bibitem{maris:etal:2021}
M.~{Maris}, E.~{Romelli}, M.~{Tomasi}, A.~{Gregorio}, M.~{Sandri},
  S.~{Galeotta} et~al., \emph{{Revised planet brightness temperatures using the
  Planck/LFI 2018 data release}},
  \href{https://doi.org/10.1051/0004-6361/202037788}{\emph{\aap} {\bfseries
  647} (2021) A104} [\href{https://arxiv.org/abs/2012.04504}{{\ttfamily
  2012.04504}}].

\bibitem{wmap:planets}
J.L.~{Weiland}, N.~{Odegard}, R.S.~{Hill}, E.~{Wollack}, G.~{Hinshaw},
  M.R.~{Greason} et~al., \emph{{Seven-year Wilkinson Microwave Anisotropy Probe
  (WMAP) Observations: Planets and Celestial Calibration Sources}},
  \href{https://doi.org/10.1088/0067-0049/192/2/19}{\emph{\apjs} {\bfseries
  192} (2011) 19} [\href{https://arxiv.org/abs/1001.4731}{{\ttfamily
  1001.4731}}].

\bibitem{Liebe:1993}
C.~Liebe, \emph{Pattern recognition of star constellations for spacecraft
  applications}, \href{https://doi.org/10.1109/62.180383}{\emph{IEEE Aerospace
  and Electronic Systems Magazine} {\bfseries 8} (1993) 31}.

\bibitem{Liebe:2002}
C.C.~{Liebe}, \emph{{Accuracy performance of star trackers - a tutorial}},
  \href{https://doi.org/10.1109/TAES.2002.1008988}{\emph{IEEE Transactions on
  Aerospace Electronic Systems} {\bfseries 38} (2002) 587}.

\bibitem{howell:2006}
S.B.~{Howell}, \emph{{Handbook of CCD Astronomy}}, vol.~5 of \emph{Cambridge
  Observing Handbooks for Research Astronomers}, Cambridge University Press,
  2~ed. (2006),
  \href{https://doi.org/10.1017/CBO9780511807909}{10.1017/CBO9780511807909}.

\bibitem{Lang:etal:2010}
D.~{Lang}, D.W.~{Hogg}, K.~{Mierle}, M.~{Blanton} and S.~{Roweis},
  \emph{{Astrometry.net: Blind Astrometric Calibration of Arbitrary
  Astronomical Images}},
  \href{https://doi.org/10.1088/0004-6256/139/5/1782}{\emph{\aj} {\bfseries
  139} (2010) 1782} [\href{https://arxiv.org/abs/0910.2233}{{\ttfamily
  0910.2233}}].

\bibitem{Schwarz:2015}
T.~Schwarz, \emph{Prototyping of a star tracker for pico-satellites},  Master's
  thesis, Massachusetts Institute of Technology, Lule\aa University of
  Technology Department of Computer Science, Electrical and Space Engineering,
  Sweden, 2015.

\bibitem{HMF:2019}
M.~Hashemi, K.M.~Mashhadi and M.~Fiuzy, \emph{Modification and hardware
  implementation of star tracker algorithms},
  \href{https://doi.org/10.1007/s42452-019-1530-0}{\emph{SN Applied Sciences}
  {\bfseries 1} (2019) }.

\bibitem{IzGh:2019}
M.~{Izadmehr} and M.K.~{Ghomi}, \emph{{Design and construction of a portable
  high resolution positioner using star patterns}},
  \href{https://doi.org/10.1007/s10509-019-3565-5}{\emph{\apss} {\bfseries 364}
  (2019) 75}.

\bibitem{arduino:cookbook}
M.~Margolis, \emph{{Arduino Cookbook}}, O'Reily (Mar., 2011).

\bibitem{avr:programming}
E.~Williams, \emph{{Make: AVR Programming}}, Maker Media Inc. (Feb., 2014).

\bibitem{Abrami:Cester:1959}
A.~{Abrami} and B.~{Cester}, \emph{{Osservazioni fotometriche della Nova
  ricorrente RS Ophiuchi}}, {\emph{\memsai} {\bfseries 30} (1959) 183}.

\bibitem{drone:paonessa:etal:ska}
F.~{Paonessa}, L.~Ciorba, G.~Virone, P.~Bolli, A.~Magro, A.~McPhail et~al.,
  \emph{{SKA-Low Prototypes Deployed in Australia: Synoptic of the UAV-Based
  Experimental Results}},
  \href{https://doi.org/doi.org/10.46620/20-0021}{\emph{URSI Radio Science
  Letters} {\bfseries 2} (2020) 1}.

\bibitem{drone:ciorba:etal}
L.~Ciorba, G.~Virone, F.~Paonessa, M.~Righero, E.~De~Lera~Acedo and
  S.~Matteoli, \emph{{Large Horizontal Near-Field Scanner Based on a
  Non-Tethered Unmanned Aerial Vehicle}},
  \href{https://doi.org/10.1109/OJAP.2022.3173741}{\emph{IEEE Open Journal of
  Antennas and Propagation} {\bfseries 3} (2022) 568}.

\bibitem{drone:virone:etal:lofarA}
G.~Virone, F.~Paonessa, L.~Ciorba, S.~Matteoli, P.~Bolli, S.~Wijnholds et~al.,
  \emph{{Measurement of the LOFAR-HBA beam patterns using an unmanned aerial
  vehicle in the near field}},
  \href{https://doi.org/10.1117/1.JATIS.8.1.011005}{\emph{J. Astron. Telesc.
  Instrum. Syst.} {\bfseries 8} (2023) 011005}.

\bibitem{drone:dininni:etal:lofarB}
P.~Di~Ninni, P.~Bolli, F.~Paonessa, G.~Pupillo, G.~Virone and S.J.~Wijnholds,
  \emph{{Electromagnetic Analysis and Experimental Validation of the LOFAR
  Radiation Patterns}},
  \href{https://doi.org/10.1155/2019/9191580}{\emph{International Journal of
  Antennas and Propagation} (2019) }.

\bibitem{drone:paonessa:etal:2020}
F.~Paonessa, G.~Virone, L.~Ciorba, G.~Addamo, M.~Lumia and G.~Dassano,
  \emph{{Design and Verification of a Q-Band Test Source for UAV-Based
  Radiation Pattern Measurements}},
  \href{https://doi.org/10.1109/TIM.2020.3031127}{\emph{IEEE Transactions on
  Instrumentation and Measurement} {\bfseries 69} (2020) 9366}.

\bibitem{report:drone:test:2018}
M.~{Maris}, S.~{Sartor}, F.~{Paonessa}, O.~{Peverini},  and E.~Y.,
  \emph{Lspe-strip: Str - uav geometric intercalibration demonstration test},
  Tech. Rep. LSPE-STRIP-RP-019, Issue/Rev. 1.0, The STRIP Collaboration
  (December, 2018).

\bibitem{astropy:2013}
{Astropy Collaboration}, T.P.~{Robitaille}, E.J.~{Tollerud}, P.~{Greenfield},
  M.~{Droettboom}, E.~{Bray} et~al., \emph{{Astropy: A community Python package
  for astronomy}},
  \href{https://doi.org/10.1051/0004-6361/201322068}{\emph{\aap} {\bfseries
  558} (2013) A33} [\href{https://arxiv.org/abs/1307.6212}{{\ttfamily
  1307.6212}}].

\bibitem{astropy:2018}
A.M.~{Price-Whelan}, B.M.~{Sip{\H{o}}cz}, H.M.~{G{\"u}nther}, P.L.~{Lim},
  S.M.~{Crawford}, S.~{Conseil} et~al., \emph{{The Astropy Project: Building an
  Open-science Project and Status of the v2.0 Core Package}},
  \href{https://doi.org/10.3847/1538-3881/aabc4f}{\emph{\aj} {\bfseries 156}
  (2018) 123}.

\bibitem{asymptote:2010}
J.C.~Bowman, \emph{{Asymptote: Interactive \TeX-aware 3D vector graphics}},
  {\emph{TUGBOAT: The Communications of the \TeX\ Users Group} {\bfseries 31:2}
  (2010) 203}.

\bibitem{emcee}
D.~{Foreman-Mackey}, D.W.~{Hogg}, D.~{Lang} and J.~{Goodman}, \emph{{emcee: The
  MCMC Hammer}}, \href{https://doi.org/10.1086/670067}{\emph{\pasp} {\bfseries
  125} (2013) 306} [\href{https://arxiv.org/abs/1202.3665}{{\ttfamily
  1202.3665}}].

\bibitem{Healpy}
A.~{Zonca}, L.~{Singer}, D.~{Lenz}, M.~{Reinecke}, C.~{Rosset}, E.~{Hivon}
  et~al., \emph{{healpy: equal area pixelization and spherical harmonics
  transforms for data on the sphere in Python}},
  \href{https://doi.org/10.21105/joss.01298}{\emph{The Journal of Open Source
  Software} {\bfseries 4} (2019) 1298}.

\bibitem{PER-GRA:2007}
F.~P\'erez and B.E.~Granger, \emph{{IP}ython: a system for interactive
  scientific computing},
  \href{https://doi.org/10.1109/MCSE.2007.53}{\emph{Computing in Science and
  Engineering} {\bfseries 9} (2007) 21}.

\bibitem{matplotlib}
J.D.~{Hunter}, \emph{Matplotlib: A 2d graphics environment},
  \href{https://doi.org/10.1109/MCSE.2007.55}{\emph{Computing in Science and
  Engineering} {\bfseries 9} (2007) 90}.

\bibitem{2020SciPy-NMeth}
P.~{Virtanen}, R.~{Gommers}, T.E.~{Oliphant}, M.~{Haberland}, T.~{Reddy},
  D.~{Cournapeau} et~al., \emph{{SciPy 1.0: Fundamental Algorithms for
  Scientific Computing in Python}},
  \href{https://doi.org/https://doi.org/10.1038/s41592-019-0686-2}{\emph{Nature
  Methods} (2020) }.

\bibitem{mckinney-proc-scipy-2010}
W.~McKinney, \emph{Data structures for statistical computing in python},  in
  \emph{Proceedings of the 9th Python in Science Conference}, S.~van~der Walt
  and J.~Millman, eds., pp.~51 -- 56, 2010.

\end{thebibliography}\endgroup

\end{document}